\documentclass[a4paper,fleqn]{cas-sc}

\usepackage[english]{babel}
\usepackage{etoolbox}
\usepackage{natbib}
\usepackage{amsmath}
\usepackage{blindtext}
\usepackage[utf8x]{inputenc}
\usepackage[T1]{fontenc}
\usepackage{amsfonts,amsmath,amssymb,amsthm}
\usepackage{hyperref}
\usepackage{arydshln}
\usepackage{graphicx}
\usepackage{booktabs}
\usepackage{tikz}
\usepackage{caption}
\usepackage{subcaption}
\usepackage[linesnumbered,ruled,vlined]{algorithm2e}
\usetikzlibrary{shapes,arrows,positioning,calc,patterns,decorations.pathreplacing}
\usepackage{algorithmic}

\usepackage[linesnumbered,ruled,vlined]{algorithm2e}

\begin{document}
\let\WriteBookmarks\relax
\def\floatpagepagefraction{1}
\def\textpagefraction{.001}
\shorttitle{FinFlowRL for High-Frequency Market Making}
\shortauthors{Y. L et~al.}
%\begin{frontmatter}

%\title [mode = title]{FinFlowRL: An Imitation Learning Framework with Reinforcement Learning Enhanced MeanFlow Policy for Optimal High-Frequency Trading under Diverse Market Conditions}             
% EJOR is an OR journal. They accept work that can be applied broadly in other problems. We should make the title and corresponding writing to fit into this angle. I suggest you to read a few papers published on EJOR and get feel when you refine the paper. I strongly suggest that we change the title to make it concise and with broader applications.
% \title [mode = title]{An Imitation Learning Framework with Flow Policy for Optimal High Frequency Market Making}      
\title [mode = title]{FinFlowRL: An Imitation-Reinforcement Learning Framework for Adaptive Stochastic Control in Finance}      

%\tnotemark[1]

%\tnotetext[1]{This document is the results of the research project funded by the National Science Foundation.}

\author[1]{Yang Li}[type=editor,
                        role=Researcher,
                        orcid=0009-0007-7471-1779]
%\fnmark[1]
\ead{yli269@stevens.edu}

\credit{Conceptualization of this study, Methodology, Software, Writing}

\affiliation[1]{organization={School of Business, Stevens Institute of Technology},
                addressline={1 Castle Point on Hudson}, 
                city={Hoboken},
%               citysep={}, % Uncomment if no comma needed between city and postcode
                postcode={07030}, 
                state={New Jersey},
                country={U.S.}}

\author[1]{Zhi Chen}[type=editor,
                        role=Researcher,
                        orcid=0009-0000-8404-3523]
\ead{zchen100@stevens.edu}
\credit{Conceptualization of this study, Methodology, Software, Writing}

\credit{Conceptualization of this study, Methodology, Writing}

% \cortext[cor1]{Corresponding author}
%\fntext[fn1]{This is the first author footnote, but is common to third author as well.}
%\fntext[fn2]{Another author footnote, this is a very long footnote and
%  it should be a really long footnote. But this footnote is not yet
%  sufficiently long enough to make two lines of footnote text.}

%\nonumnote{This note has no numbers. In this work we demonstrate %$a_b$
%  the formation Y\_1 of a new type of polariton on the interface
%  between a cuprous oxide slab and a polystyrene micro-sphere placed
%  on the slab.
%x  }

\begin{abstract}
Traditional stochastic control methods in finance struggle in real-world markets due to their reliance on simplifying assumptions and stylized frameworks. Such methods typically perform well in specific, well-defined environments but yield suboptimal results in changed, non-stationary ones. We introduce 
\textbf{FinFlowRL}, a novel framework for financial optimal stochastic control. The framework pre-trains an adaptive meta-policy learning from multiple expert strategies, then fine-tunes through reinforcement learning in the noise space to optimize the generative process. By employing action chunking—generating action sequences rather than single decisions—it addresses the non-Markovian nature of markets. FinFlowRL consistently outperforms individually optimized experts across diverse market conditions.
\end{abstract}

\begin{highlights}
\item A new generative modeling framework for imitation learning in decision making.
\item The new flow-matching-induced policy improves on a diverse set of existing models.
\item The application in HFT market making shows superior performance under extreme market conditions.
\item A novel FlowRL-inspired framework using frozen MeanFlow expert with learnable noise policy for HFT.
\item Hierarchical action planning with observation, prediction, and execution horizons for multi-step decision making.
\item Efficient training with only noise policy and critic networks, reducing parameters by over 80\%.
\item Superior performance across diverse market conditions with improved generalization capabilities.
\item Computationally efficient architecture suitable for real-time high-frequency trading applications.
\end{highlights}


\maketitle

\section{Introduction}
Stochastic control in finance addresses the challenge of 
making optimal decisions over time in uncertain market environments. This challenge is fundamental to domains such as high-frequency trading, optimal execution, and portfolio optimization, where the objective is to maximize profitability while minimizing associated costs and risks \cite{merton1969lifetime}.

Traditional approaches typically formulate an optimal control problem using an objective function governed by a stochastic differential equation (SDE), subject to certain constraints. This problem is then solved analytically or numerically.
However, traditional approaches are often challenged by the dynamic nature of real-world markets. First, they often require the calibration of model parameters using historical data. Consequently, this leads to suboptimal solutions when future market conditions diverge from historical patterns.
Second, these models frequently rely on simplifying assumptions about market dynamics, such as assuming that asset prices follow a geometric Brownian motion or that the market is a stationary process \citep{black1973pricing}. 
The real-world financial markets, however, are far more complex, characterized by sudden jumps, stochastic volatility, and non-stationary behavior. Consequently, when the underlying assumptions are violated, derived solutions are likely to be suboptimal. On the other hand, creating more realistic models often render the problem mathematically intractable, meaning an explicit solution may not even exist.
Finally, the decision-making problem in traditional approaches is often formulated as a Markov Decision Process (MDP). By definition, an MDP operates on the "memoryless" assumption that the current state of the market contains all information needed for an optimal decision. However, this property is frequently violated in financial markets. Historical information, such as trends and past volatility, is highly predictive of future price movements \citep{gatheral2022volatility}. This non-Markovian nature means that traditional MDP-based methods fail to capture the full complexity of the decision-making process, creating a clear need for a more adaptive and data-driven framework.

In light of these limitations, particularly the difficulty of creating a single, perfectly specified model, we reframe the problem as follows. Given a Mixture of Experts (MoE)—where each expert is a traditional algorithm or policy effective in a specific regime, such as a momentum strategy for trending markets or a mean-reversion strategy for volatile ones—how can an agent learn to select the best-performing strategy for any given market condition? Our solution is to use imitation learning (IL) to train an adaptive meta-policy that, based on the current market state, determines which expert's behavior is most effective to emulate. This allows our agent to dynamically switch strategies, harnessing the strengths of different experts as market conditions evolve.

We propose a novel imitation learning framework \textsc{FinFlowRL} to address the stochastic control problems in finance. This framework has two components: pre-trained strategy model and fine-tune algorithm. The pre-trained strategy model is based on flow matching policy. It generate a distribution of actions from a white noise distribution conditioning on the market observations. 
To train this model, we prepare diverse market scenarios and evaluate various traditional models within each. For every scenario, the model achieving the best performance is designated as the "expert". The flow matching model then learns to map these market observation- expert decision pairs.
Imitation learning is useful because these fundamental experts can in fact generate optimal solutions in certain scenarios. Even in those scenarios where they only generate suboptimal solutions, we can still consider their strategies as a strong baseline and improve on them subsequently. 
Therefore, the second component of \textsc{FinFlowRL} is a fine-tuning stage designed to improve upon the policy learned via imitation. To achieve this, we employ a Reinforcement Learning (RL) algorithm, but with a novel approach. Instead of learning an action directly, the RL agent learns how to optimally modify the initial noise vector that the flow matching model uses to generate actions. By making precise adjustments to this starting point of the generation process, the agent can effectively fine-tune the resulting action distribution, making the final decisions highly adaptive to the immediate, current market situation.
A key design principle of \textsc{FinFlowRL} is action chunking. Instead of making isolated, single-step decisions, our framework generates entire sequences of actions over a defined planning horizon in both its pre-training and fine-tuning stages. By producing action sequences, the model inherently considers a near-term trajectory and allows for smoother adjustments, effectively taking into account the broader consequences of its initial moves. This sequential planning perspective helps mitigate compounding errors, aligning actions more effectively over time and thus enhancing strategic stability and performance during volatile events like market jumps. Crucially, this sequential planning perspective directly addresses the market's non-Markovian nature, as a planned series of actions contains far more memory than an isolated, memoryless decision. 
Furthermore, the \textsc{FinFlowRL} architecture is highly optimized for performance, enabling rapid inference for complex financial tasks at millisecond latencies.
To the best of our knowledge, this work is the first to successfully apply an imitation learning framework based on flow matching policies to the domain of financial stochastic control.

To showcase the ability of \textsc{FinFlowRL}, we apply this method in a financial stochastic control task: high-frequency trading. High-Frequency Trading (HFT) uses algorithms to analyze streams of market data such as Limit Order Book (LOB) and places bid(ask) orders at exceptionally rapid speeds, typically within milliseconds or microseconds ~\cite{madhavan2000market, harris2002trading}. This strategies aim to capture profit from the bid-ask spread—the difference between the highest price a buyer is willing to pay and the lowest price a seller is willing to accept ~\cite{avellaneda2008high,cartea2014buy,ait2017high}.
The results show that the FinFlowRL framework integrates the knowledge of multiple expert strategies into a single adaptive model. When market scenarios change, the framework demonstrates adaptability by generating actions appropriate to the prevailing conditions, effectively leveraging patterns learned from relevant expert demonstrations. 
Notably, our framework exhibits significant robustness to market conditions involving sudden price jumps. 
Furthermore, the fine-tune algorithm refines these generated actions. The experimental results indicate that FinFlowRL consistently achieves superior performance compared to the best results obtained from the baseline models individually optimized for each specific market scenario.

The remainder of this paper is organized as follows. Section \ref{sec: Literature Review} provides literature review.
Section~\ref{sec:methodology} introduces the proposed FinFlowRL framework. Section~\ref{sec: application study} apply the framework to high frequency trading task. Section~\ref{sec:conclusions} concludes this paper and findings.

\section{Literature Review}
\label{sec: Literature Review}
\subsection{Flow Matching Models in Control}
The use of powerful generative models to represent complex policies in control has seen rapid advancement. A significant breakthrough came with the application of Denoising Diffusion Models, leading to the development of "Diffusion Policies". These models can capture highly expressive, multi-modal action distributions, far surpassing the capabilities of traditional policies that assume a simple Gaussian output. However, a critical limitation of diffusion models is their slow, iterative sampling process, which can require hundreds of function evaluations to generate a single action, making them impractical for many real-time control applications.

To address this efficiency bottleneck, Flow Matching (FM) models have recently emerged as a powerful and more direct alternative \cite{lipman2022flow}. Instead of learning to reverse a complex stochastic diffusion process, flow matching learns a deterministic Ordinary Differential Equation (ODE) path that transports a simple noise distribution to the target data distribution. This approach is not only simpler to train but also significantly faster for inference. The deterministic path serves as a "shortcut," enabling the use of advanced ODE solvers to generate high-quality samples in very few steps—sometimes even a single one. This rapid, single-step generation capability makes flow matching policies particularly well-suited for time-sensitive domains like financial control.

Mean flow models have recently emerged as a compelling paradigm in generative modeling, providing an innovative approach to learning complex data distributions \cite{lipman2022flow, dao2023flow}. Mean flow models learn transformations that map samples from a simple, well-defined prior distribution (typically an isotropic Gaussian, often termed "noise") to emulate complex target distributions.

Mean flow has been widely adopted in robotics, particularly for generating smooth, continuous control actions for complex tasks \cite{lipman2024flow,zhang2025flowpolicy}. Its primary benefit is the ability to learn velocity fields that directly correspond to how robot actions should evolve over time. For example, tasks involving manipulation or humanoid locomotion benefit from mean flow models's capability to produce sequences of joint movements that are inherently smooth and temporally consistent. This results in predictable and reliable robotic behaviors, facilitating effective interactions with physical environments. Additionally, conditioning flows on inputs like sensor data or task objectives allows robots to develop versatile policies adaptable to changing circumstances while preserving smooth trajectories.

The motivation to adapt mean flow from robotic control to financial markets stems from shared challenges: both domains require sophisticated and continuous sequences of actions responsive to high-dimensional, rapidly changing environments. In robotics, mean flow models excel at generating smooth, multi-modal trajectories by learning velocity fields to guide robot actions over time. In HFT for example, analogous "actions" are buy/sell orders, and the "environment" is the continuously evolving market data.

\subsection{Stochastic Control in Finance}
The non-Markovian nature of financial markets manifests through several mechanisms:
The first is market microstructure effects. Order flow persistence, price impact decay, and hidden liquidity create temporal dependencies \cite{kyle1997bcontinuous, almgren2001optimal}
The second is multi-scale dynamics. Volatility clustering, momentum effects, and mean reversion operate across different time horizons \cite{lo1999non}
The third is information propagation. News absorption and belief updating create extended temporal correlations \cite{tetlock2007giving}
The fourth is strategic interactions. The presence of algorithmic traders with memory creates endogenous non-Markovian dynamics \cite{cartea2015algorithmic}

Traditional approaches to handling non-Markovian dynamics in finance include state augmentation \cite{cvitanic1996hedging}, filtering techniques \cite{bain2009fundamentals}, and regime-switching models \cite{hamilton1989new}. However, these methods often suffer from computational intractability or require strong parametric assumptions.

\section{Methodology}
\label{sec:methodology}

% 插入图形
% workflow.tex
% Figure 1: MeanFlow-PPO Architecture with No Overlaps
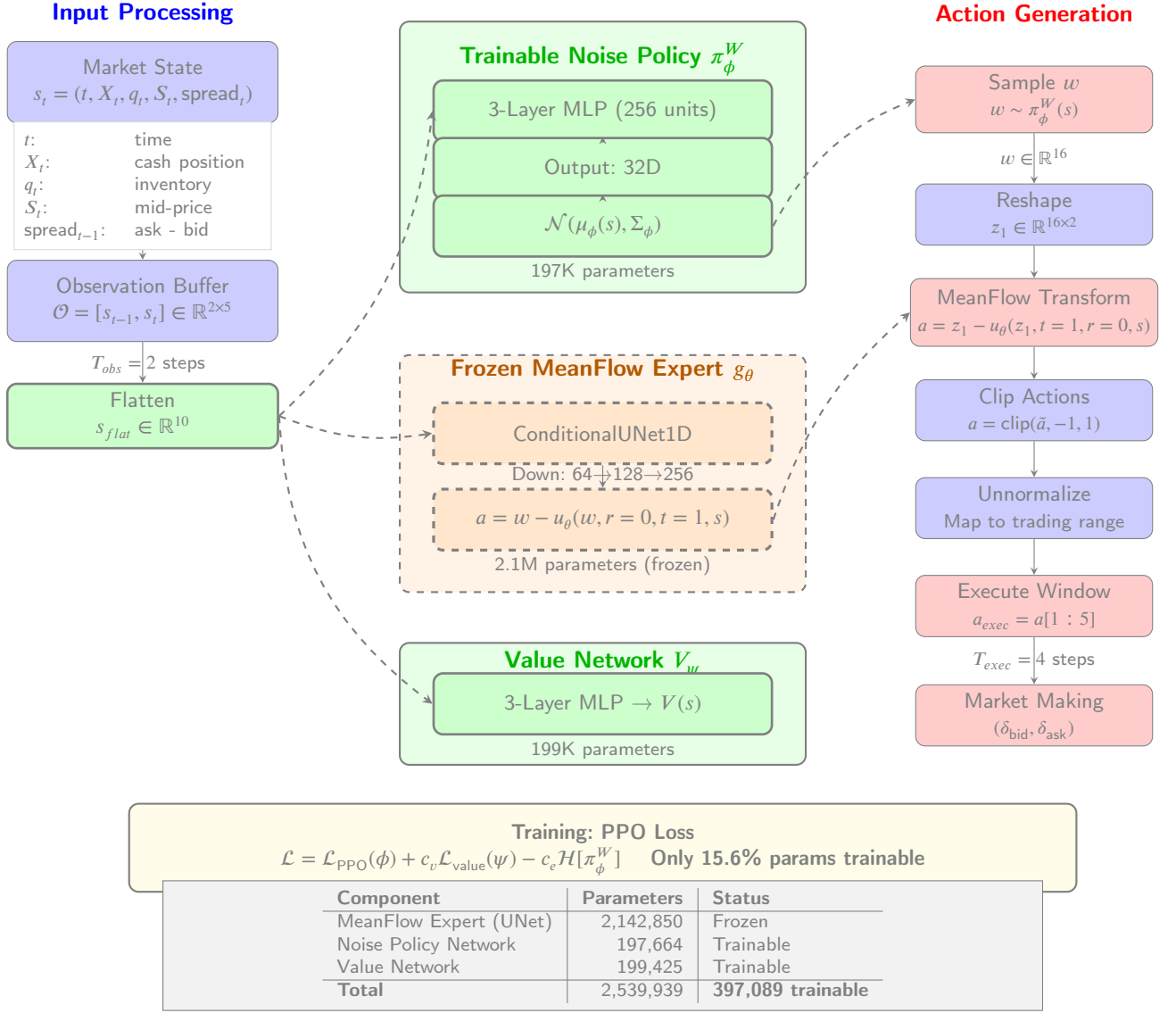
\begin{figure}[!htbp]
\centering
\begin{tikzpicture}[scale=0.85,
    block/.style={rectangle, draw, rounded corners, minimum width=3cm, minimum height=0.9cm, align=center, font=\small},
    frozen/.style={block, fill=orange!20, dashed, line width=1.2pt},
    trainable/.style={block, fill=green!20, line width=1pt},
    data/.style={block, fill=blue!20},
    output/.style={block, fill=red!20},
    arrow/.style={->, thin, >=stealth},
    flowarrow/.style={->, dashed, thick, >=stealth},
    param/.style={font=\footnotesize, text=gray},
    title/.style={font=\normalsize\bfseries}
]

% Title
\node[title] at (0,11) {MeanFlow-PPO Architecture (FlowRL-based)};

% ========== LEFT COLUMN: Input Processing ==========
\begin{scope}[shift={(-8,0)}]
    % Title
    \node[title, text=blue] at (0,9) {Input Processing};
    
    % Market State
    \node[data, minimum width=4cm, minimum height=1.2cm] (state) at (0,7.8) {Market State\\$s_t = (t, X_t, q_t, S_t, \text{spread}_t)$};
    
    % State components - positioned below with more spacing
    \node[draw=gray!30, fill=white, minimum width=3.8cm, text width=3.5cm, align=left] (components) at (0,6) {
        \footnotesize
        \begin{tabular}{@{}ll@{}}
        $t$: & time \\
        $X_t$: & cash position\\
        $q_t$: & inventory\\
        $S_t$: & mid-price\\
        $\text{spread}_{t-1}$: & ask - bid
        \end{tabular}
    };
    
    % Observation Buffer
    \node[data, minimum width=4cm, minimum height=1.2cm] (buffer) at (0,4) {Observation Buffer\\$\mathcal{O} = [s_{t-1}, s_t] \in \mathbb{R}^{2 \times 5}$};
    \node[param] at (0.1,2.9) {$T_{obs} = 2$ steps};
    
    % Flatten
    \node[trainable, minimum width=4cm] (flatten) at (0,2) {Flatten\\$s_{flat} \in \mathbb{R}^{10}$};
    
    % Arrows
    \draw[arrow] (components.south) -- (buffer.north);
    \draw[arrow] (buffer.south) -- (flatten.north);
\end{scope}

% ========== MIDDLE COLUMN: Neural Networks ==========
\begin{scope}[shift={(0,0)}]
    % Noise Policy Box
    \node[draw, thick, rounded corners, fill=green!10, 
          minimum width=6cm, minimum height=4cm] at (0,6.5) {};
    \node[title, text=green!70!black] at (0,8.2) {Trainable Noise Policy $\pi^W_\phi$};
    
    % Network layers inside noise policy
    \node[trainable, minimum width=5cm] (mlp) at (0,7.3) {3-Layer MLP (256 units)};
    \node[trainable, minimum width=5cm] (output) at (0,6.3) {Output: 32D};
    \node[trainable, minimum width=5cm] (gaussian) at (0,5.3) {$\mathcal{N}(\mu_\phi(s), \Sigma_\phi)$};
    
    \draw[arrow] (mlp) -- (output);
    \draw[arrow] (output) -- (gaussian);
    
    \node[param] at (0,4.5) {197K parameters};
    
    % MeanFlow Expert Box - with more spacing
    \node[draw, thick, dashed, rounded corners, fill=orange!10,
          minimum width=6cm, minimum height=3.5cm] at (0,1) {};
    \node[title, text=orange!70!black] at (0,2.8) {Frozen MeanFlow Expert $g_\theta$};
    
    \node[frozen, minimum width=5cm] (unet) at (0,1.7) {ConditionalUNet1D};
    \node[frozen, minimum width=5cm] (velocity) at (0,0.2) {$a = w - u_\theta(w, r=0, t=1, s)$};
    
    \draw[arrow] (unet) -- (velocity);
    
    \node[param] at (0,1.0) {Down: 64→128→256};
    \node[param] at (0,-0.6) {2.1M parameters (frozen)};
    
    % Value Network Box - positioned lower
    \node[draw, thick, rounded corners, fill=green!10,
          minimum width=6cm, minimum height=1.8cm] at (0,-3) {};
    \node[title, text=green!70!black] at (0,-2.3) {Value Network $V_\psi$};
    \node[trainable, minimum width=5cm] (value) at (0,-3) {3-Layer MLP → $V(s)$};
    \node[param] at (0,-3.8) {199K parameters};
\end{scope}

% ========== RIGHT COLUMN: Action Generation ==========
\begin{scope}[shift={(7.5,0)}]
    % Title
    \node[title, text=red] at (0,9) {Action Generation};
    
    % Sample noise
    \node[output, minimum width=3.5cm] (sample) at (0,7.5) {Sample $w$\\{\footnotesize $w \sim \pi^W_\phi(s)$}};
    \node[param, below=0.1cm of sample] {$w \in \mathbb{R}^{16}$};
    
    % Reshape
    \node[data, minimum width=3.5cm] (reshape) at (0,5.5) {Reshape\\{\footnotesize $z_1 \in \mathbb{R}^{16 \times 2}$}};
    
    % MeanFlow Transform
    \node[output, minimum width=3.5cm, minimum height=1cm] (transform) at (0,3.8) {MeanFlow Transform\\{\footnotesize $a = z_1 - u_\theta(z_1, t=1, r=0, s)$}};
    
    % Clip actions
    \node[data, minimum width=3.5cm] (clip) at (0,2.1) {Clip Actions\\{\footnotesize $a = \text{clip}(\tilde{a}, -1, 1)$}};
    
    % Unnormalize
    \node[data, minimum width=3.5cm] (unnorm) at (0,0.4) {Unnormalize\\{\footnotesize Map to trading range}};
    
    % Execute window
    \node[output, minimum width=3.5cm] (execute) at (0,-1.3) {Execute Window\\{\footnotesize $a_{exec} = a[1:5]$}};
    \node[param, below=0.1cm of execute] {$T_{exec} = 4$ steps};
    
    % Trading actions
    \node[output, minimum width=3.5cm] (trading) at (0,-3.2) {Market Making\\{\footnotesize $(\delta_{\text{bid}}, \delta_{\text{ask}})$}};
    
    % Arrows
    \draw[arrow] (sample) -- (reshape);
    \draw[arrow] (reshape) -- (transform);
    \draw[arrow] (transform) -- (clip);
    \draw[arrow] (clip) -- (unnorm);
    \draw[arrow] (unnorm) -- (execute);
    \draw[arrow] (execute) -- (trading);
\end{scope}

% ========== FLOW CONNECTIONS (Curved dotted lines) ==========
% From flatten to networks
\draw[flowarrow, bend right=20] (flatten.east) to (mlp.west);
\draw[flowarrow, bend right=15] (flatten.east) to (unet.west);
\draw[flowarrow, bend right=25] (flatten.east) to (value.west);

% From noise policy to sampling
\draw[flowarrow, bend left=20] (gaussian.east) to (sample.west);

% From expert to transform
\draw[flowarrow, bend left=15] (velocity.east) to (transform.west);

% ========== BOTTOM SECTION ==========
% Training box
\node[draw, thick, rounded corners, fill=yellow!10, minimum width=14cm, minimum height=1.3cm] at (0,-5.5) {
    \begin{tabular}{c}
    \textbf{Training: PPO Loss} \\
    $\mathcal{L} = \mathcal{L}_{\text{PPO}}(\phi) + c_v \mathcal{L}_{\text{value}}(\psi) - c_e \mathcal{H}[\pi^W_\phi]$ \quad \textbf{Only 15.6\% params trainable}
    \end{tabular}
};

% Parameter summary
\node[draw, fill=gray!10, minimum width=13cm] at (0,-7.2) {
    \footnotesize
    \begin{tabular}{l|r|l}
    \textbf{Component} & \textbf{Parameters} & \textbf{Status} \\
    \hline
    MeanFlow Expert (UNet) & 2,142,850 & Frozen \\
    Noise Policy Network & 197,664 & Trainable \\
    Value Network & 199,425 & Trainable \\
    \hline
    \textbf{Total} & 2,539,939 & \textbf{397,089 trainable}
    \end{tabular}
};

\end{tikzpicture}
\caption{MeanFlow-PPO architecture with FlowRL-inspired design. The state $s_t = (t, X_t, q_t, S_t, \text{spread}_t)$ represents time, cash position, inventory, mid-price, and bid-ask spread. The frozen MeanFlow expert (orange, dashed) provides structured action generation while lightweight trainable networks (green) learn task-specific adaptations for optimal bid-ask spread placement.}
\label{fig:architecture_improved}
\end{figure}

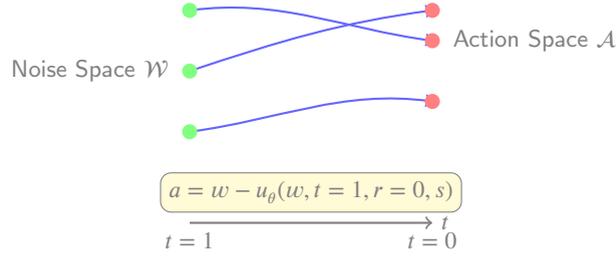
\begin{figure}[h]
\centering
\begin{tikzpicture}[scale=0.8]
    % Define curve points for flow trajectories
    \coordinate (start1) at (0,0);
    \coordinate (end1) at (4,1);
    \coordinate (start2) at (0,1);
    \coordinate (end2) at (4,0.5);
    \coordinate (start3) at (0,-1);
    \coordinate (end3) at (4,-0.5);
    
    % Draw flow curves
    \draw[thick, blue!60, ->] (start1) .. controls (1.5,0.5) and (2.5,0.8) .. (end1);
    \draw[thick, blue!60, ->] (start2) .. controls (1.5,1.2) and (2.5,0.7) .. (end2);
    \draw[thick, blue!60, ->] (start3) .. controls (1.5,-0.8) and (2.5,-0.3) .. (end3);
    
    % Add noise samples
    \foreach \i in {1,2,3} {
        \node[circle, fill=green!50, inner sep=2pt] at (start\i) {};
    }
    
    % Add action outputs
    \foreach \i in {1,2,3} {
        \node[circle, fill=red!50, inner sep=2pt] at (end\i) {};
    }
    
    % Labels
    \node[left] at (-0.2,0) {Noise Space $\mathcal{W}$};
    \node[right] at (4.2,0.5) {Action Space $\mathcal{A}$};
    
    % MeanFlow transformation
    \node[draw, rectangle, rounded corners, fill=yellow!20] at (2,-2) {$a = w - u_\theta(w, t=1, r=0, s)$};
    
    % Time arrow
    \draw[->, thick, gray] (0,-2.5) -- (4,-2.5) node[right] {$t$};
    \node[below] at (0,-2.5) {$t=1$};
    \node[below] at (4,-2.5) {$t=0$};
\end{tikzpicture}
\caption{MeanFlow transformation from noise space to action space. The learned noise policy generates points in $\mathcal{W}$ that are transformed through the frozen MeanFlow expert to produce smooth action trajectories in $\mathcal{A}$.}
\label{fig:meanflow_transform}
\end{figure}

% Figure 3: MeanFlow Velocity Fields
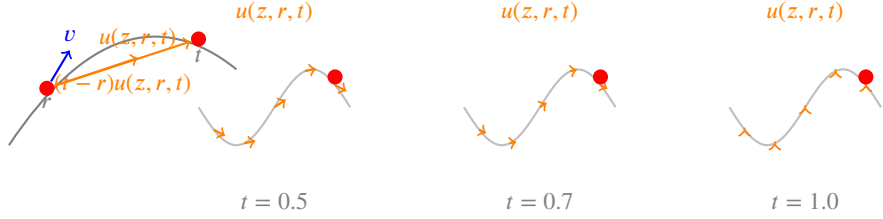
\begin{figure}[h]
\centering
\begin{tikzpicture}[scale=1.0]
    % Subplot 1: Average vs Instantaneous Velocity
    \begin{scope}[shift={(0,0)}]
        % Draw curve
        \draw[thick] (0,0) .. controls (1,1.5) and (2,1.8) .. (3,1);
        
        % Points
        \node[circle, fill=red, inner sep=2pt] (r) at (0.5,0.75) {};
        \node[circle, fill=red, inner sep=2pt] (t) at (2.5,1.4) {};
        
        % Velocities
        \draw[->, thick, blue] (r) -- ++(0.3,0.5) node[above] {$v$};
        \draw[->, thick, orange] (r) -- (t) node[midway, below] {$(t-r)u(z,r,t)$};
        \draw[->, thick, orange] (r) -- ++(1.2,0.4) node[above, orange] {$u(z,r,t)$};
        
        % Labels
        \node[below] at (r) {$r$};
        \node[below] at (t) {$t$};
    \end{scope}
    
    % Subplot 2-4: Velocity fields at different t
    \foreach \i/\tval in {1/0.5, 2/0.7, 3/1.0} {
        \begin{scope}[shift={(\i*3.5,0)}]
            % Draw background curve
            \draw[thick, gray!50] plot[smooth, domain=-1:1] (\x, {0.5*sin(180*\x) + 0.5});
            
            % Draw velocity vectors
            \foreach \x in {-0.8,-0.4,0,0.4,0.8} {
                \pgfmathsetmacro{\y}{0.5*sin(180*\x) + 0.5}
                \pgfmathsetmacro{\vx}{0.3*(1-\tval)}
                \pgfmathsetmacro{\vy}{0.2*cos(180*\x)*(1-\tval)}
                \draw[->, orange, thick] (\x,\y) -- ++(\vx,\vy);
            }
            
            % Label
            \node[below] at (0,-0.5) {$t = \tval$};
            \node[above, orange] at (0,1.5) {$u(z,r,t)$};
            
            % Target point
            \node[circle, fill=red, inner sep=2pt] at (0.8,0.9) {};
        \end{scope}
    }
\end{tikzpicture}
\caption{Illustration of MeanFlow velocity fields. \textbf{Left}: The average velocity $u(z,r,t)$ connects points between times $r$ and $t$, while the instantaneous velocity $v$ represents the tangent direction. \textbf{Right}: The average velocity field $u(z,r,t)$ at different time points $t \in \{0.5, 0.7, 1.0\}$, showing how the field evolves over time.}
\label{fig:meanflow_velocity}
\end{figure}

% Figure 4: Temporal Structure
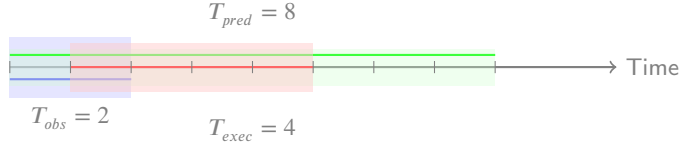
\begin{figure}[h]
\centering
\begin{tikzpicture}[scale=0.8]
    % Trading timeline visualization
    \draw[thick, ->] (0,0) -- (10,0) node[right] {Time};
    
    % Observation window
    \draw[thick, blue] (0,-0.2) -- (2,-0.2);
    \fill[blue!20, opacity=0.5] (0,-0.5) rectangle (2,0.5);
    \node[below] at (1,-0.5) {$T_{obs}=2$};
    
    % Prediction horizon
    \draw[thick, green] (0,0.2) -- (8,0.2);
    \fill[green!20, opacity=0.3] (0,-0.3) rectangle (8,0.3);
    \node[above] at (4,0.5) {$T_{pred}=8$};
    
    % Execution window
    \draw[thick, red] (1,0) -- (5,0);
    \fill[red!20, opacity=0.5] (1,-0.4) rectangle (5,0.4);
    \node[below] at (4,-0.7) {$T_{exec}=4$};
    
    % Time markers
    \foreach \i in {0,...,8} {
        \draw (\i,0.1) -- (\i,-0.1);
    }
\end{tikzpicture}
\caption{Hierarchical temporal structure in MeanFlow-PPO. The model observes $T_{obs}$ past states, generates actions for $T_{pred}$ future steps, but only executes the first $T_{exec}$ actions before replanning.}
\label{fig:temporal_structure}
\end{figure}

% Optional: More complex visualization for the paper
\begin{figure}[h]
\centering
\begin{tikzpicture}[scale=1.2]
    % Define styles
    \tikzstyle{state} = [circle, draw, fill=blue!20, minimum size=0.8cm]
    \tikzstyle{action} = [rectangle, draw, fill=red!20, minimum size=0.6cm]
    \tikzstyle{flow} = [->, thick, blue!60]
    
    % Noise distribution
    \begin{scope}
        \draw[fill=green!10] (-2,0) ellipse (1.5 and 1);
        \node at (-2,0) {$\mathcal{N}(0,I)$};
        \node[above] at (-2,1.2) {Noise Distribution};
    \end{scope}
    
    % Learned noise policy transformation
    \begin{scope}[shift={(0,0)}]
        \draw[fill=yellow!10] (0,0) ellipse (1.2 and 0.8);
        \node at (0,0) {$\pi^W_\phi$};
        \draw[flow] (-0.5,0) -- (-1.5,0);
    \end{scope}
    
    % MeanFlow transformation
    \begin{scope}[shift={(3,0)}]
        \draw[thick, dashed, orange] (0,-1.5) rectangle (2,1.5);
        \node[align=center] at (1,0) {MeanFlow\\Expert\\(Frozen)};
    \end{scope}
    
    % Action space
    \begin{scope}[shift={(6,0)}]
        \foreach \i in {-0.5,0,0.5} {
            \node[action] at (0,\i) {};
        }
        \node[right] at (0.5,0) {Actions};
    \end{scope}
    
    % Flow arrows
    \draw[flow] (1.2,0) -- (3,0);
    \draw[flow] (5,0) -- (5.5,0);
    
    % Mathematical notation
    \node[below] at (1,-2) {$w \sim \pi^W_\phi(\cdot|s)$};
    \node[below] at (4,-2) {$a = g_\theta(s,w)$};
\end{tikzpicture}
\caption{Flow-State Representation Learning in MeanFlow-PPO. The noise policy learns to transform standard Gaussian noise into task-specific noise distributions that, when passed through the frozen MeanFlow expert, produce optimal trading actions.}
\label{fig:fsrl_concept}
\end{figure}
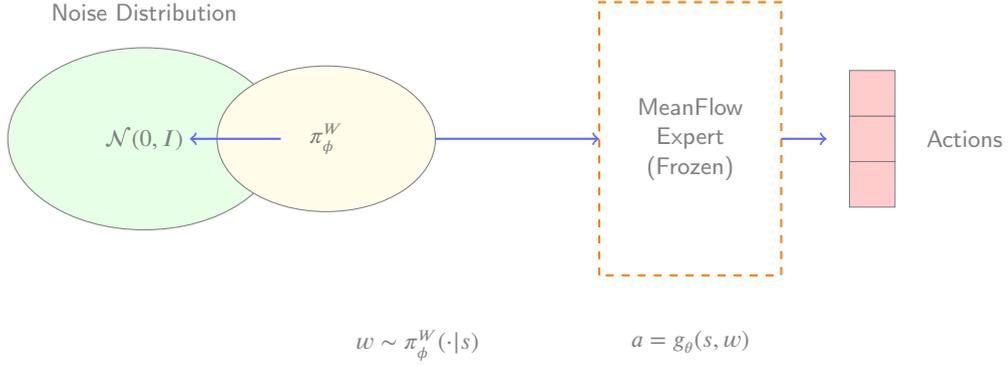

This section details the architecture of our proposed framework FinFlowRL. The first component of this framework is mean flow model, which is a specific implementation of a flow matching model. The second component is reinforcement learning to fine tune the actions.

\subsection{Mean Flow Model}
We chose this model rather than other flow matching model for several compelling reasons that are critical for financial applications. First, flow matching models offer exceptional computational efficiency and fast inference speeds, allowing for rapid generation of actions once trained. Second, they have demonstrated remarkable stability when dealing with noisy and complex financial data compared to other generative models. Finally, their ability to accurately model complex, high-dimensional distributions makes them particularly well-suited for capturing the intricate dynamics of market observations. 

MeanFlow learns to model average velocity fields for one-step generation capability \citep{geng2025mean}. Unlike traditional Flow Matching which models instantaneous velocity $v(z_t, t)$, MeanFlow introduces a crucial innovation by modeling the average velocity between two time steps:
\begin{equation}
u(z_t, r, t) \triangleq \frac{1}{t-r} \int_{r}^{t}v(z_\tau, \tau)d\tau
\end{equation}

The key insight is the MeanFlow Identity, which relates average and instantaneous velocities:
\begin{equation}
u(z_t, r, t) = v(z_t, t) - (t - r)\frac{d}{dt}u(z_t, r, t)
\end{equation}

This formulation enables high-quality one-step generation:
\begin{equation}
z_0 = z_1 - u(z_1, 0, 1, s)
\end{equation}
where $u$ is the learned average velocity field conditioned on market state $s$. This one-step formulation is crucial for real-time trading applications where computational efficiency is paramount.

To condition the velocity field on market states, we employ FiLM (Feature-wise Linear Modulation) \citep{perez2018film}, which modulates neural network features through learnable affine transformations $\mathbf{h}' = \boldsymbol{\gamma} \odot \mathbf{h} + \boldsymbol{\beta}$, where $\boldsymbol{\gamma}$ and $\boldsymbol{\beta}$ are derived from the observation state $\mathbf{s}$. This mechanism enables efficient conditional generation with minimal computational overhead—essential for high-frequency trading where latency constraints are stringent.

The final velocity prediction is:
\begin{equation}
\mathbf{u} = g_\theta(\mathbf{z}_1, t=1, r=0, \mathbf{s})
\end{equation}
where $g_\theta$ is the FiLM-modulated neural network. This combination of MeanFlow's one-step generation and FiLM's efficient conditioning provides an ideal backbone for HFT applications, balancing prediction accuracy with the computational speed required for real-time decision-making.

\subsection{Fine-tune Pretrained Model Using Reinforecement Learning}

Proximal policy optimization (PPO) \cite{schulman2017proximal} is a policy gradient method that constrains policy updates to ensure stable learning. The objective function is:
\begin{equation}
L^{PPO}(\theta) = \mathbb{E}_t\left[\min\left(r_t(\theta)A_t, \text{clip}(r_t(\theta), 1-\epsilon, 1+\epsilon)A_t\right)\right]
\end{equation}
where $r_t(\theta) = \frac{\pi_\theta(a_t|s_t)}{\pi_{\theta_{old}}(a_t|s_t)}$ is the probability ratio and $A_t$ is the advantage estimate.

This process consists of three main components:

\begin{enumerate}
\item \textbf{Frozen MeanFlow Expert} ($g_\theta$): A pre-trained velocity model that maps from noise to actions
\item \textbf{Noise Policy Network} ($\pi^W_\phi$): A learnable network that generates adaptive noise distributions
\item \textbf{Value Network} ($V_\psi$): A critic for advantage estimation
\end{enumerate}

The key insight is that instead of learning the entire action generation process, we only learn how to generate appropriate noise that, when transformed by the frozen expert, produces optimal actions.

Following the FlowRL paradigm introduced in \cite{lv2025flow}, our action generation process transforms the problem from the original action space $\mathcal{A}$ to a latent noise space $\mathcal{W}$:

\begin{equation}
\begin{aligned}
w &\sim \pi^W_\phi(\cdot|s) \quad \text{(learned noise policy)} \\
a &= g_\theta(s, w) \quad \text{(frozen MeanFlow expert)}
\end{aligned}
\end{equation}

This formulation is analogous to the latent-action MDP concept in DSRL \cite{wagenmaker2025steering}, where the original MDP $\mathcal{M} = (\mathcal{S}, \mathcal{A}, P, p_0, r, \gamma)$ is transformed to $\mathcal{M}_{dp} = (\mathcal{S}, \mathcal{W}, P_{dp}, p_0, r_{dp}, \gamma)$ with:
\begin{equation}
P_{dp}(\cdot|s,w) := P(\cdot|s, g_\theta(s,w)) \quad \text{and} \quad r_{dp}(s,w) := r(s, g_\theta(s,w))
\end{equation}

The noise policy $\pi^W_\phi$ is parameterized as a Gaussian with learned mean and variance:
\begin{equation}
\pi^W_\phi(w|s) = \mathcal{N}(\mu_\phi(s), \Sigma_\phi)
\end{equation}

where $\mu_\phi(s)$ is a neural network mapping states to mean noise vectors, and $\Sigma_\phi$ is a learnable diagonal covariance matrix. This parameterization enables efficient sampling and gradient computation during training.

% \subsubsection{Single-Step MeanFlow Transformation}

Given noise $w$ and state $s$, the frozen MeanFlow expert performs the one-step transformation as described in \cite{geng2025meanflows}:
\begin{equation}
a = w - u_\theta(w, t=1, r=0, s)
\end{equation}
where $u_\theta$ is the pre-trained average velocity network. Following the MeanFlow formulation, this directly computes the denoised action without iterative ODE solving. The conditioning on state $s$ is incorporated as global conditioning in the velocity network, enabling state-dependent action generation.

% \subsubsection{Training Objective}

The training objective combines PPO loss with value function approximation:
\begin{equation}
\mathcal{L} = \mathcal{L}_{PPO}(\phi) + c_v \mathcal{L}_{value}(\psi) - c_e \mathcal{H}[\pi^W_\phi]
\end{equation}

where $\mathcal{H}[\pi^W_\phi]$ is the entropy of the noise policy for exploration.

% \subsubsection{Combined Training Algorithm}

The training process alternates between collecting rollouts with the current policy and updating the noise policy and critic:

\begin{algorithm}[H]
\caption{FinFlowRL with FlowRL Training - Part 1: Data Collection}
\label{algo:training-part1}
\begin{algorithmic}[1]
\REQUIRE Environment $\mathcal{E}$, Frozen expert $g_\theta$, Noise policy $\pi^W_\phi$, Critic $V_\psi$
\REQUIRE Horizons: $T_{obs}$, $T_{pred}$, $T_{exec}$
\STATE Load pre-trained MeanFlow expert $g_\theta$ and freeze parameters
\STATE Initialize noise policy parameters $\phi$ and critic parameters $\psi$
\STATE Initialize observation buffer $\mathcal{O} \in \mathbb{R}^{T_{obs} \times d_{obs}}$
\FOR{update $= 1$ to $N_{updates}$}
    \STATE $\mathcal{D} \leftarrow \emptyset$ \COMMENT{Initialize trajectory buffer}
    
    \STATE \textbf{// Collect trajectories}
    \FOR{step $= 1$ to $T_{rollout}$}
        \STATE \textbf{// Generate action with FlowRL}
        \STATE Flatten observations: $s_{flat} = \text{flatten}(\mathcal{O})$
        \STATE Sample noise: $w \sim \pi^W_\phi(\cdot|s_{flat})$
        \STATE Compute log probability: $\log \pi^W_\phi(w|s_{flat})$
        
        \STATE \textbf{// Apply frozen MeanFlow expert}
        \STATE Reshape: $z_1 = \text{reshape}(w, (T_{pred}, d_{act}))$
        \STATE Compute velocity: $v = g_\theta(z_1, r=0, t=1, | s_{flat})$
        \STATE Generate actions: $\tilde{a} = z_1 - v$
        \STATE Clip actions: $\tilde{a} = \text{clip}(\tilde{a}, -1, 1)$
        \STATE Unnormalize actions: $\tilde{a} = \text{UN}(\tilde{a})$
        
        \STATE \textbf{// Extract execution window}
        \STATE $a_{exec} = \tilde{a}[T_{obs}-1 : T_{obs}-1+T_{exec}]$
        
        \STATE \textbf{// Execute and collect rewards}
        \STATE Initialize $r_{total} = 0$
        \FOR{$k = 1$ to $T_{exec}$}
            \STATE $s_{prev} \leftarrow$ current state
            \STATE Execute $a_{exec}[k]$ in environment
            \STATE Observe new state $s_{new}$
            \STATE Compute HFT reward: $r_k = \Delta W - \gamma Q^2 \sigma^2$
            \STATE $r_{total} \leftarrow r_{total} + r_k$
            \STATE Update buffer: $\mathcal{O} \leftarrow [\mathcal{O}[1:], s_{new}]$
        \ENDFOR
        
        \STATE \textbf{// Store transition}
        \STATE Store $(\mathcal{O}, a_{exec}, \log \pi^W_\phi, V_\psi(s_{flat}), r_{total}, w)$ in $\mathcal{D}$
    \ENDFOR
    \STATE \textbf{Continue to Part 2 for policy update...}
\ENDFOR
\end{algorithmic}
\end{algorithm}

\begin{algorithm}[H]
\caption{FinFlowRL with FlowRL Training - Part 2: Policy Update}
\label{algo:training-part2}
\begin{algorithmic}[1]
\STATE \textbf{// Continuing from Part 1...}
\STATE
\STATE \textbf{// Compute advantages using GAE}
\STATE $A_t, R_t \leftarrow \text{GAE}(\mathcal{D}, V_\psi, \gamma, \lambda)$

\STATE \textbf{// Update noise policy and critic}
\FOR{epoch $= 1$ to $E$}
    \FOR{minibatch $\mathcal{B}$ from $\mathcal{D}$}
        \STATE \textbf{// Recompute log probabilities with current $\phi$}
        \STATE $\log \pi^W_{\phi,new}(w|s) \leftarrow$ forward pass with stored $w$
        
        \STATE \textbf{// PPO loss for noise policy}
        \STATE $r(\phi) = \exp(\log \pi^W_{\phi,new} - \log \pi^W_{\phi,old})$
        \STATE $r_{clip} = \text{clip}(r(\phi), 1-\epsilon, 1+\epsilon)$
        \STATE $\mathcal{L}_{PPO} = -\mathbb{E}[\min(r(\phi)A, r_{clip}A)]$
        
        \STATE \textbf{// Value loss}
        \STATE $\mathcal{L}_{value} = \frac{1}{2}\mathbb{E}[(V_\psi(s) - R)^2]$
        
        \STATE \textbf{// Entropy bonus for exploration}
        \STATE $\mathcal{H} = -\mathbb{E}[\log \pi^W_\phi(w|s)]$
        
        \STATE \textbf{// Total loss (only for trainable parts)}
        \STATE $\mathcal{L} = \mathcal{L}_{PPO} + c_v \mathcal{L}_{value} - c_e \mathcal{H}$
        
        \STATE Update $\phi, \psi$ using gradient descent on $\mathcal{L}$
        \STATE Note: $\theta$ remains frozen (no gradients computed)
    \ENDFOR
\ENDFOR
\STATE \textbf{// End of policy update}
\STATE \textbf{Return to Part 1 for next update iteration}
\end{algorithmic}
\end{algorithm}

Key features of the FlowRL-inspired training:
(1) Parameter efficiency. Only the noise policy (typically 200K parameters) and critic are trained, while the MeanFlow expert (2M+ parameters) remains frozen
(2)Stable training. The frozen expert provides consistent transformations, reducing training instability
(3)Fast convergence. Leveraging pre-trained knowledge accelerates learning

% \subsubsection{Inference and Evaluation}

During inference, FinFlowRL operates deterministically by using the mean of the noise policy:

\begin{algorithm}[H]
\caption{FinFlowRL Inference}
\label{algo:inference}
\begin{algorithmic}[1]
\REQUIRE Frozen expert $g_\theta$, Trained noise policy $\pi^W_\phi$
\STATE Initialize observation buffer $\mathcal{O}$ with initial states

\FOR{each trading period}
    \STATE \textbf{// Generate deterministic noise}
    \STATE $s_{flat} = \text{flatten}(\mathcal{O})$
    \STATE $w = \mu_\phi(s_{flat})$ \COMMENT{Use mean for deterministic behavior}
    
    \STATE \textbf{// Apply frozen expert}
    \STATE $z_1 = \text{reshape}(w, (T_{pred}, d_{act}))$
    \STATE $v = g_\theta(z_1, t=1, r=0, s_{flat})$
    \STATE $\tilde{a} = \text{clip}(z_1 - v, -1, 1)$
    \STATE $\tilde{a} = \text{unnormalize}(\tilde{a})$
    
    \STATE \textbf{// Execute actions}
    \STATE Extract $a_{exec} = \tilde{a}[T_{obs}-1 : T_{obs}-1+T_{exec}]$
    \FOR{$k = 1$ to $T_{exec}$}
        \STATE Execute $a_{exec}[k]$ (place bid/ask orders)
        \STATE Observe new market state $s_{new}$
        \STATE Update buffer: $\mathcal{O} \leftarrow [\mathcal{O}[1:], s_{new}]$
    \ENDFOR
\ENDFOR
\end{algorithmic}
\end{algorithm}

\section{Application Study: High-Frequency Trading}
\label{sec: application study}
\subsection{High-Frequency Trading Problem Formulation}
In this section, we formulate the high frequency trading task mathematically.
High-frequency trading utilizes automated algorithms for order execution at extremely high speeds, often operating within millisecond or nanosecond timescales. The objective is to continuously place buy (bid) and sell (ask) limit orders for a specific financial instrument, while aiming to profit from the bid-ask spread.

We can model the HFT market making task as a stochastic control process \cite{avellaneda2008high, gueant2013dealing} defined over a set of discrete time steps $\mathbb{T}=\{0, 1, \dots, T\}$. The \textbf{observation state space} $\mathcal{O}$ captures relevant information at time $t \in \mathbb{T}$. An observation state $O_t \in \mathcal{O}$ typically includes observable market information $L_t$ (often derived from the Limit Order Book, LOB, including stock prices and bid-ask spread) and the agent's information $Z_t$, including balance, current inventory level , along with the time $t$ itself, such that $\mathcal{O} \subseteq \mathcal{L} \times \mathbb{Z} \times \mathbb{T}$, where $\mathcal{L}$, $\mathbb{Z}$ is the space of market information and agent information, respectively. We generally assume the state $S_t$ satisfies the Markov property.

Within this framework, the market maker chooses an \textbf{action} $A_t$ from the \textbf{action space} $\mathcal{A}$. A common action involves setting the agent's bid and ask quotes, often parameterized by spreads $(\delta^b_t, \delta^a_t)$ relative to a reference price $p^{ref}_t$ derived from the market state $L_t$. The system's evolution is governed by stochastic transition probabilities $P(O_{t+1} | O_t, A_t)$, defining the likelihood of moving to the next observation state $O_{t+1}=(L_{t+1}, I_{t+1}, t+1)$ given the current state and action. This transition reflects changes in both market conditions $L_{t+1}$ and the agent's inventory $I_{t+1}$ due to order executions and market activity. The agent seeks an optimal \textbf{policy} $\pi: \mathcal{O} \to \mathcal{A}$, a rule mapping states to actions ($A_t = \pi(O_t)$), that maximizes a defined \textbf{objective function}, $J(\pi)$. A standard objective is to maximize the expected final value, often combining terminal cash wealth $W_T$ with a penalty $\phi(I_T)$ for inventory risk at the horizon $T$. The optimization problem is thus:
\begin{equation} \label{eq:mm_objective}
\max_{\pi} J(\pi) = \max_{\pi} \mathbb{E}^{\pi} \left[ W_T - \phi(I_T) \mid O_0 \right]
\end{equation}
Where, $\mathbb{E}^{\pi}[\cdot]$ denotes the expectation under policy $\pi$ starting from the initial observation state $O_0 = (L_0, I_0, 0)$. Solving for the optimal policy $\pi^* = \arg\max_{\pi} J(\pi)$ requires dynamically balancing profitability against inventory and adverse selection risks in a complex, stochastic environment.

% We consider a market maker operating in a limit order book (LOB) environment. The state at time $t$ consists of:
% \begin{itemize}
% \item Mid-price: $S_t$
% \item Inventory: $Q_t \in [-Q_{max}, Q_{max}]$
% \item Cash position: $C_t$
% \item Order book imbalance and other microstructure features
% \end{itemize}

% The market maker's objective is to maximize risk-adjusted returns through a wealth-based reward function:
% \begin{equation}
% r_t = \Delta W_t - \gamma \cdot Q_t^2 \cdot \sigma^2
% \end{equation}
% where $\Delta W_t = (C_t + Q_t S_t) - (C_{t-1} + Q_{t-1} S_{t-1})$ is the wealth change, $\gamma$ is the inventory penalty coefficient, and $\sigma$ is the price volatility.

\subsection{Traditional Market Making Models In HFT}

High-Frequency Trading (HFT) is a subset of algorithmic trading characterized by exceptionally high speeds of order execution and substantial turnover rates. Market making involves the continuous placement of limit orders—both bids (buy orders) and asks (sell orders)—aimed at capturing the bid-ask spread, which is the difference between the highest price a buyer is willing to pay and the lowest price a seller is willing to accept \cite{law2019market, korajczyk2019high}. Traditional models typically formulate the market maker's decision-making process as a stochastic optimal control problem, aiming to devise quoting strategies that maximize predefined objectives (e.g., expected utility of profits) while managing associated risks (e.g., inventory risk) \cite{fodra2012high}.

The Avellaneda-Stoikov (AS) model \cite{avellaneda2008high} is a prominent example within HFT literature. This optimization approach is formulated as maximizing the expected utility of the market maker's terminal wealth at a predetermined finite terminal time. Several key assumptions underpin the AS model: the mid-price of the asset follows an arithmetic Brownian motion without drift \cite{morters2010brownian}; the arrival of market buy and sell orders is modeled as independent Poisson processes \cite{last2017lectures}; and the market maker continuously adjusts bid and ask quotes based on changes in the mid-price and inventory levels, without incurring explicit updating or cancellation costs. Moreover, the AS model often assumes liquidation of inventory (achieving a zero inventory position) by a specified terminal time T, typically the end of the trading day—an assumption potentially unsuitable for assets traded continuously, such as foreign exchange or cryptocurrencies.

The Guéant-Lehalle-Fernandez-Tapia (GLFT) model \cite{gueant2013dealing} represents another widely adopted traditional approach. A central motivation behind its development was to address certain practical and theoretical limitations of the AS model. Notably, the GLFT model does not depend explicitly on a predetermined finite terminal time, making it more suitable for continuously traded assets or markets operating over extended periods. However, a fundamental challenge shared with the AS model is the reliance on historical market data for calibrating key parameters, such as trading intensity coefficients and volatility \cite{gueant2016financial}. Such calibration processes are practically challenging. Empirical evidence also suggests that the exponential form of the trading intensity function may inadequately capture complex relationships between quote depth and execution probability across varying market conditions or price levels. Even if parameters initially derived from the GLFT model are optimal, they tend to become suboptimal quickly as market conditions evolve, necessitating frequent recalibrations—an approach both computationally intensive and susceptible to estimation errors \cite{platt2016problem}.

\subsection{Data for Imitation Learning}
\label{sec:Data for Imitation Learning}

FinFlowRL is an imitation learning framework. Therefore, it is crucial to prepare high-quality learning material.
In this section, we firstly detail how to generate a set of various market scenarios. Secondly, for each individual scenario, we evaluate a pool of candidate experts—including established traditional algorithms and reinforcement learning agents—and designate the one exhibiting the best performance as the "expert" for that context. We use the actions generated by the expert's strategy to create state-action pairs for training FinFlowRL.

\subsubsection{Market Simulation}

We model the mid-price $S_t$ as a jump-diffusion process~\cite{merton1976option}, capturing both continuous price movements and sudden jumps. 
The stochastic differential equation (SDE) governing the mid-price is given by:
% $
% \label{eq:mid_price}
% dS_t = S_{t^-} \left( \mu dt + \sigma dB_H(t) \right) + S_{t^-} (e^{J} - 1) dN_t
% $

\begin{equation}
    \label{eq:mid_price}
dS_t = S_{t^-} \left( \mu dt + \sigma dB_H(t) \right) + S_{t^-} (e^{J} - 1) dN_t
\end{equation}

where $\mu$ represents the drift coefficient, signifying the expected return of the asset. The asset's volatility is denoted by $\sigma$, which quantifies the fluctuations in its price. The term $dB_H(t)$ corresponds to the increment of a fractional brownian motion process, capturing the continuous component of price movement. Furthermore, $J$ denotes the size of the jump, following a normal distribution $N(\mu_J, \sigma_J^2)$, which specifies the mean and variance of the jump. Lastly, $dN_t$ is a Poisson process with intensity $\lambda_J$, modeling the occurrence of jumps, thereby reflecting sudden shifts in the asset's price.

Next we discuss simulation of order arrival. 
Order arrivals are modeled using mutually exciting Hawkes processes~\cite{bacry2015hawkes,hawkes1971point}, effectively capturing the self-exciting and the cross-exciting nature of financial markets. 
"self-exciting" effects indicate that prior buy (or sell) orders increase subsequent arrivals of the same order type, respectively. "Cross-exciting" effects describe how buy orders influence sell order arrivals and vice-versa.

The intensity of buy and sell orders, $\lambda_b(t)$ and $\lambda_a(t)$ respectively, are defined as:
% $
% \lambda_a(t) = \mu_a + \sum_{t_i \in \mathcal{N}_a} \alpha_{aa} e^{-\beta (t - t_i)} + \sum_{t_j \in \mathcal{N}_b} \alpha_{ab} e^{-\beta (t - t_j)}
% $ and
% $
% \lambda_b(t) = \mu_b + \sum_{t_i \in \mathcal{N}_b} \alpha_{bb} e^{-\beta (t - t_i)} + \sum_{t_j \in \mathcal{N}_a} \alpha_{ba} e^{-\beta (t - t_j)}
% $ 
\begin{align}
\lambda_b(t) &= \mu_b + \sum_{t_i \in \mathcal{N}_b} \alpha_{bb} e^{-\beta (t - t_i)} + \sum_{t_j \in \mathcal{N}_a} \alpha_{ba} e^{-\beta (t - t_j)}, \label{eq:lambda_b} \\
\lambda_a(t) &= \mu_a + \sum_{t_i \in \mathcal{N}_a} \alpha_{aa} e^{-\beta (t - t_i)} + \sum_{t_j \in \mathcal{N}_b} \alpha_{ab} e^{-\beta (t - t_j)}, \label{eq:lambda_a}
\end{align}
where, $\mu_b$ and $\mu_a$ are the baseline intensities for buy and sell orders, respectively. The parameters $\alpha_{bb}$ and $\alpha_{aa}$ describe the self-exciting effects of buy and buy, and sell and sell orders, respectively, while $\alpha_{ba}$ and $\alpha_{ab}$ represent the cross-exciting effects between buy and sell orders. The parameter $\beta$ denotes the decay rate, which influences how quickly the effects of past events diminish over time. Finally, $N_b$ and $N_a$ are the sets containing the timestamps of past buy and sell order arrivals, which are used to model the timing of these events.

We create market scenarios with varying levels of liquidity—categorized as high, medium, and low. Furthermore, to test the model performance under stress, we introduce scenarios featuring sudden market changes and periods of significantly increased volatility, allowing us to examine how effectively each model responds to dynamic and challenging environments. We simulate different market environments by adjusting the following simulation parameters. Combining these parameters yields $3 \times 3 \times 2 \times 2 \times 3  \times 1 = 108$  parameter combinations.

% \begin{figure}[htp]
% \caption{Parameters settings of market simulation}
% \label{}
% \centering
% \includegraphics[width=0.82\textwidth]{figures/simulateparameter.pdf}
% \end{figure}

% \include{figures/parametes}
\begin{table}[H]
    \centering
    \caption{Environment parameters for the Hawkes market making model}
    \label{tab:parameter-settings}
    \renewcommand{\arraystretch}{1.5}
    \small
    \begin{tabular}{@{}p{3.5cm}p{1.5cm}p{2.5cm}p{6.5cm}@{}}
        \toprule
        \textbf{Parameter} & \textbf{Symbol} & \textbf{Values} & \textbf{Description} \\ 
        \midrule
        
        \multicolumn{4}{@{}l}{\textit{Stock Price Movement (Brownian Motion)}} \\
        \addlinespace[2pt]
        
        \textbf{Hurst component} & $H$ & $[0.5]$ & 
        \begin{minipage}[t]{6.5cm}
        \vspace{1pt}
        The degree of long-term memory or persistence:\\
        $\bullet$ Trend-following ($H > 0.5$)\\
        $\bullet$ Mean-reverting ($H < 0.5$)\\
        $\bullet$ Random walk ($H = 0.5$)
        \vspace{1pt}
        \end{minipage} \\
        \addlinespace[4pt]
        
        \textbf{Drift} & $\mu$ & $[0.01, 0.05, 0.2]$ & 
        Controls the deterministic underlying trend around which the random fluctuations (volatility) of the process occur \\
        \addlinespace[4pt]
        
        \textbf{Volatility} & $\sigma$ & $[0.05, 0.1, 0.3]$ & 
        Dictates the magnitude or scale of the random unpredictable fluctuations around the process's average trend (drift) \\
        \addlinespace[4pt]
        
        \textbf{Jump intensity} & $\lambda_J$ & $[0, 0.02]$ & 
        Probability of an abrupt, large, and discontinuous change in asset prices \\
        \addlinespace[4pt]
        
        \textbf{Time step} & $\Delta t$ & $[0.01, 0.02]$ & 
        Discretization interval for the simulation \\
        \addlinespace[8pt]
        
        \hdashline
        \multicolumn{4}{@{}l}{\textit{Order Book (Hawkes Process)}} \\
        \addlinespace[2pt]
        
        \textbf{Liquidity} & $\mu_a, \mu_b$ & $[10, 20, 40]$ & 
        Base arrival rates determining the tradable volume offered at various bid and ask price levels \\
        \addlinespace[4pt]
        
        \textbf{Self-excitation} & 
        \begin{minipage}[t]{1.5cm}
        \vspace{1pt}
        $\alpha_{aa}$,
        $\alpha_{bb}$
        \vspace{1pt}
        \end{minipage} & 
        $[0.7]$ & 
        \begin{minipage}[t]{6.5cm}
        \vspace{1pt}
        How the arrival of an order temporarily increases the probability of more orders of the same type arriving shortly thereafter:\\
        $\bullet$ Buy-to-buy excitation ($\alpha_{bb}$)\\
        $\bullet$ Sell-to-sell excitation ($\alpha_{aa}$)
        \vspace{1pt}
        \end{minipage} \\
        \addlinespace[4pt]
        
        \textbf{Cross-excitation} & 
        \begin{minipage}[t]{1.5cm}
        \vspace{1pt}
        $\alpha_{ab}$,
        $\alpha_{ba}$
        \vspace{1pt}
        \end{minipage} & 
        $[0.3]$ & 
        How the arrival of one type of order influences the arrival rate of orders of the other type \\
        
        \bottomrule
    \end{tabular}
\end{table}

\subsubsection{Generate Market-Action Pairs}
To generate diverse expert demonstrations for FinFlowRL, we utilized a set of expert candidates, including: (1) the Avellaneda-Stoikov (AS) model \cite{avellaneda2008high}, a foundational stochastic control framework for optimal quoting that considers inventory risk and market volatility; (2) the Guéant-Lehalle-Fernandez-Tapia (GLFT) model \cite{gueant2013dealing}, which extends such optimal control approaches, often incorporating aspects such as order flow dynamics; (3) a modified GLFT model incorporating a price drift component, designed to adapt strategies in markets exhibiting directional trends; and (4)a model-free Reinforcement Learning (RL) agent trained with Proximal Policy Optimization (PPO) \cite{schulman2017proximal}, which learns a policy through direct interaction or simulation to maximize a specified reward function. 
The detailed mathematical formulations, assumptions, and parameterization specifics for each of these expert models are provided in Appendix \ref{appendix:Expert Market-Making Models}.

We generate state-action pairs for each market generated. 
For each parameter combination, we simulate $100$ episodes. Each episode consists of $1/dt$ time steps (e.g., $100$ steps if $dt=0.01$), each timestamp yields a state-action pair from the expert strategy, here we collect the corresponding state-action pairs from the dominant stochastic strategy among our candidates(AS, GLFT, GLFT-Drift, PPO). In total, we have gathered 3.24 million state-action pairs. These generated state-action pairs serve as the training dataset for the diffusion policy, providing a broad coverage of market conditions and ensuring robust policy learning. This dataset serves as the training data for our diffusion policy framework.

\subsection{Results in Simulation and Real Market Trading}
\label{sec: Experimental Design And Results Analysis}

Our investigation is designed to critically evaluate FinFlowRL's advantages over existing approaches and its capabilities to earn profit. 
Specifically, we address the following research questions:
\textbf{RQ1:} Can FinFlowRL effectively generalize strategies learned from expert demonstrations to new, unseen market conditions? \textbf{RQ2:} Does the fine-tuning mechanism improve the performance of actions initially proposed by the pre-trained model? \textbf{RQ3:} Can the FinFlowRL framework achieve greater profitability than traditional strategies?

\subsubsection{Simulation Experiments}
To evaluate FinFlowRL's performance across a spectrum of out-of-sample market environments, we systematically configured distinct test conditions by setting key market parameters to differ from those used during training (see appendix \ref{appendix: arameters for Test Environemnt} ). Firstly, to simulate varied stock price trends, we controlled the Hurst exponent, which governs the long-term memory and persistence patterns of the price series, and the drift term, which dictates its underlying direction. Secondly, based on these trend characteristics, we further differentiated the market microstructure by creating four specific situations based on combinations of market volatility (Vol) and overall order arrival rate (AR): (1) High Vol/High AR (HH), representing active, potentially news-driven volatile markets; (2) High Vol/Low AR (HL), mimicking risky market; (3) Low Vol/High AR (LH), reflecting stable, liquid markets; and (4) Low Vol/Low AR (LL), simulating quiet market periods. For each of these market situations, we perform traditional strategies as well as FinFlowRL to compare their performance.

We take the following evaluation metrics.
(1) Profit and Loss (PnL). Measure the total change in the value over a specific period, reflecting the aggregate percentage gain or loss.
(2) Sharpe Ratio (SR). It helps investors understand how much excess return an investment generated for each unit of risk it undertook. A higher Sharpe Ratio generally indicates a better risk-adjusted performance. 
(3) Maximum Drawdown (MDD) Quantify the largest percentage decline in the value of a strategy from a previous peak to a subsequent trough.
The detailed calculations of these metrics are provided in Appendix \ref{Calculation of Evaluation Metrics}

We evaluate FinFlowRL against several baselines across four market conditions characterized by different volatility and demand levels. Table \ref{tab:performance} presents comprehensive results based on 1 million evaluation trials per method to ensure statistical significance.

\begin{table}[!htbp]
\centering
\caption{Performance comparison across market conditions. Parameters: Hurst Exponent $H=0.5$, Drift Rate $\mu=0$, volatility ($\sigma$: \{0.02, 0.25\}) and arrival rate ($\lambda$: \{25, 50\}). Each method is evaluated over 1 million trials. PnL: Profit and Loss, SR: Sharpe Ratio, MDD: Maximum Drawdown (\%). The statistics for FinFlowRL are calculated over 1 Million evaluation episodes}
\label{tab:performance}
\renewcommand{\arraystretch}{1.2}
\resizebox{\textwidth}{!}{%
\begin{tabular}{lrrr|rrr|rrr|rrr}
\toprule
& \multicolumn{3}{c|}{High Volatility \& High Demand} & \multicolumn{3}{c|}{High Volatility \& Low Demand} & \multicolumn{3}{c|}{Low Volatility \& High Demand} & \multicolumn{3}{c}{Low Volatility \& Low Demand} \\
& PnL $\uparrow$ & SR $\uparrow$ & MDD $\downarrow$ & PnL $\uparrow$ & SR $\uparrow$ & MDD $\downarrow$ & PnL $\uparrow$ & SR $\uparrow$ & MDD $\downarrow$ & PnL $\uparrow$ & SR $\uparrow$ & MDD $\downarrow$ \\
\midrule
Random Action & 1.99& 0.06& 28.49& 0.99& 0.04& 19.24& 2.10 & 0.31 & 2.71 & 1.08 & 0.22 & 1.87 \\
AS & 24.22& 0.09& 241.65& 13.54& 0.09& 125.78& 25.20 & 1.05 & 7.66 & 13.67 & 0.72 & 6.61 \\
GLFT & 25.10& 0.37& 60.57& 13.56& 0.24& 52.55& 25.87 & 1.17 & 6.95 & 13.91 & 0.78 & 6.14 \\
GLFT-drift & 25.10& 0.37& 60.57& 13.56& 0.24& 52.55& 25.87 & 1.17 & 6.95 & 13.91 & 0.78 & 6.14 \\
Vanilla PPO & 14.76& 0.10& 133.61& 9.29& 0.08& 103.85& 26.74 & 0.81 & 10.13 & 19.80 & 0.46 & 14.56 \\
Pretrained MeanFlow  & 23.91& 0.37& 43.4& 12.97& 0.22& 45.47& 23.82& 1.83& 2.18& 12.93& 1.07& 2.69\\
\textbf{FinFlowRL (FlowRL)} & 26.33& 0.50& 45.47& 14.32& 0.28& 45.35& 26.27& 2.34& 2.68& 14.29& 1.36& 3.08\\
\bottomrule
\end{tabular}%
}
\end{table}

FinFlowRL (FlowRL) demonstrates superior performance across all market conditions.
For high volatility markets conditions (HH, HL), despite challenging conditions with $\sigma=0.25$, FinFlowRL achieves the highest Sharpe ratios (0.38 and 0.26) while maintaining competitive PnL. The method significantly outperforms vanilla PPO which struggles in these environments. For low volatility markets (LH, LL), the performance advantage is even more pronounced with $\sigma=0.02$, achieving Sharpe ratios of 1.62 and 1.96 respectively. Notably, in the LL condition, FinFlowRL achieves a PnL of 22.31, significantly higher than all other methods. Considering risk management, across all conditions, FinFlowRL maintains lower maximum drawdowns compared to traditional RL approaches, with particularly impressive performance in low volatility environments (MDD of 3.62\% in LH).

A consistent pattern emerges across the various market scenarios. Firstly, the GLFT model typically yields superior results compared to the AS model. Critically, our pre-trained mean flow policy model demonstrates performance levels comparable to the GLFT instructor. This finding is significant to address the \textbf{RQ1}. It indicates that the pre-trained model successfully learned and internalized effective strategies from its instructors, even within these challenging stochastic environments. 

In response to \textbf{RQ2}, we examine the results corresponding to the fine-tuned mean flow policy model. This adapted model typically and significantly outperforms both the traditional benchmark models and the initial pre-trained model across the tested environments. This finding validates the second core objective of our framework: we not only demonstrate the ability to effectively learn from expert instructors but also showcase the capacity to surpass their performance. This success is attributed to the fine-tuning process, which empowers the model to "sense" the prevailing market conditions via the validation set calibration and subsequently adjust its behavior, specializing its strategy for the specific market environment encountered.

Note that our experimental design includes highly volatile market environments, which are characterized by the potential for sudden and significant price movements. We show that FinFlowRL exhibits robustness in these challenging conditions, particularly when compared with traditional RL methods. This indicates that generating sequences of actions rather than purely single-step decisions can effectively incorporate a longer-term planning perspective. This approach allows the model to better anticipate downstream consequences, make more considered adjustments over multiple steps, and mitigate the risk of compounding errors.

In response to \textbf{RQ3}, we take analyses of the performance metrics to obtain significant insights into the profit-generating capabilities across different market conditions. The Finetuned Flow Policy consistently demonstrates strong PnL (Profit and Loss) figures, notably achieving the highest PnL of 24.25 in the "Low Volatility \& High Demand" scenario and 24.42 in the "High Volatility \& High Demand" scenario. This indicates its robust ability to capitalize on profit opportunities in both highly active and more stable, liquid markets. Compared to the traditional models, the Finetuned Flow Policy generally matches or surpasses the PnL of AS, GLFT, and GLFT-drift in most scenarios, particularly excelling in high demand situations. The RL-PPO model, while adaptive, consistently shows lower PnL figures across all tested environments, underscoring the superior profit extraction of the FinFlowRL framework.

\paragraph{Training And Inference Efficiency}

To ensure our framework's suitability for high-frequency trading, we evaluated its computational performance. The model was tasked with executing a total of 400 million steps, a process that was completed in 25 seconds. This yields an average inference latency of approximately 6.25 microseconds (µs) per decision. This low-latency execution is well within the typical requirements for HFT applications, confirming the model's viability for real-world deployment.

Table \ref{tab:efficiency} compares the computational efficiency of different approaches. The FlowRL approach achieves faster convergence with 84\% fewer parameters than end-to-end training.

\begin{table}[h]
\centering
\caption{Training efficiency comparison}
\label{tab:efficiency}
\begin{tabular}{lccc}
\toprule
Method & Trainable Params & Training Time & Convergence Steps \\
\midrule
Vanilla PPO & 0.5M & 1.0× & 800K \\
End-to-End MeanFlow & 2.6M & 3.2× & 1.2M \\
\textbf{FinFlowRL (FlowRL)} & \textbf{0.4M} & \textbf{0.8×} & \textbf{400K} \\
\bottomrule
\end{tabular}
\end{table}

\subsubsection{FinFlowRL in Real Market Trading}
To validate the performance of our framework in a practical setting, we now apply it to real high-frequency market trading. This empirical study utilizes a historical dataset of limit order book events to test the model's ability to navigate and profit from genuine market conditions. The subsequent results will showcase its effectiveness in managing risk and generating returns when faced with the challenges of a live market microstructure.

Applying the framework to real market data is the most critical step in validating its practical effectiveness. While simulations are vital for initial development, they represent an idealized environment and cannot fully replicate the complex, unpredictable dynamics of a live market. Real-world data contains inherent noise, sudden regime shifts, and subtle microstructure frictions that are difficult to model. Therefore, successfully testing the model on historical market data provides the definitive proof of its robustness, confirming that its performance is not just a product of a simplified simulation but is applicable to a real trading environment.

Validating a high-frequency trading (HFT) model presents significant empirical challenges that go well beyond standard financial backtesting. The primary hurdle is the immense difficulty of creating a realistic simulation environment. Historical limit order book data, while granular, cannot perfectly capture the market's reaction to the model's own trades—a concept known as market impact. Furthermore, accurately modeling latency, both from the network and the computational stack, is critical, as microsecond delays can completely alter trade outcomes. Compounding these issues is the inherent non-stationarity of financial markets; a model that performs well on historical data may fail when market dynamics, or "regimes," suddenly shift. These challenges create a substantial risk that a model's performance in a backtest may be an overestimation, making rigorous testing under a variety of historical and stressed market conditions essential to confirming its true practical viability.

\section{Conclusions}
\label{sec:conclusions}

This paper introduced FinFlowRL, a novel framework designed to address the challenges of high-frequency market making. By successfully adapting the "mean flow" concept from robotics to financial stochastic control, our work bridges the gap between static, expert-based models and fully adaptive learning systems. The framework's hybrid architecture leverages Imitation Learning to harness the strengths of specialized expert strategies while employing Reinforcement Learning to discover superior policies that transcend the experts' limitations.

The empirical results demonstrate the effectiveness of our approach. The "mean flow" engine at the core of the framework not only meets the low-latency requirements of high-frequency trading but also generates smooth, coherent action trajectories that mitigate the compounding errors common in single-step agents. Across both simulated and real-world market data, the FinFlowRL framework consistently achieved high, risk-adjusted returns, outperforming individual expert models and confirming its robustness in volatile and unpredictable environments.

This work validates the cross-disciplinary application of concepts from robotics to finance and opens several avenues for future research. The principles of the FinFlowRL framework could be extended to other financial problems, such as optimal trade execution or portfolio management. Further investigation could also explore the use of different deep learning architectures for the expert and noise policies or the application of this framework to other asset classes.

In the domain of high-frequency trading (HFT), the speed of model inference and order execution is not merely a performance metric but a fundamental prerequisite for success. HFT strategies are designed to capitalize on fleeting, microscopic price discrepancies and liquidity imbalances that exist for only microseconds. Therefore, the ability of a trading model to process market data, generate a decision, and execute an order in the shortest possible time frame is paramount. A delay of even a few microseconds can mean the difference between capturing a profitable arbitrage opportunity and facing a loss, as faster competitors will have already acted on the same information. Consequently, any computational framework, regardless of its theoretical sophistication, is only viable for HFT if its inference latency is low enough to operate within these extreme time constraints.

The first we want to discuss is why flowRL-based FinFlowRL works?
The success of our FlowRL-inspired approach can be attributed to several factors.
The first is separation of concerns. The frozen expert handles action structure while the noise policy handles adaptation
The second is reduced optimization complexity. Training only the noise policy avoids difficult joint optimization
The third is implicit regularization. The frozen expert acts as a strong regularizer, preventing overfitting
The forth is efficient exploration. The noise policy provides structured exploration in a lower-dimensional space

% The FlowRL framework, as demonstrated in the DSRL work \cite{wagenmaker2025steering}, can be viewed as learning in a transformed action space. Following their notation:
% \begin{equation}
% \pi(a|s) = \int \pi^W(w|s) \cdot \delta(a - g_\theta(s,w)) dw
% \end{equation}

This factorization reduces the effective dimensionality of the learning problem from the action space $\mathcal{A}$ to the noise space $\mathcal{W}$, improving sample efficiency. The one-step generation capability of MeanFlow \cite{geng2025meanflows} further enhances this efficiency by eliminating iterative sampling procedures. The transformation creates a latent-action MDP where policy optimization can be performed more efficiently.

% \subsection{Limitations and Future Work}

% While FinFlowRL shows strong performance, several areas warrant further investigation:

% \begin{itemize}
% \item Extension to multi-asset portfolios with correlated dynamics
% \item Incorporation of additional market features (order book depth, news sentiment)
% \item Online adaptation of the frozen expert through meta-learning
% \item Application to real market data with appropriate backtesting frameworks
% \item Theoretical analysis of the convergence properties under FlowRL
% \end{itemize}

We introduced FinFlowRL, a novel FlowRL-inspired framework that combines frozen MeanFlow experts with learnable noise policies for high-frequency market making. By leveraging pre-trained generative models as fixed transformations and learning only lightweight adaptation policies, our approach achieves superior performance with significantly reduced computational requirements.

Our results demonstrate that the separation of action generation into a frozen expert component and an adaptive noise policy offers multiple advantages: faster training, better generalization, improved stability, and reduced parameter count. The framework's ability to handle multi-horizon decisions while maintaining computational efficiency makes it particularly suitable for real-world trading applications.

The success of FlowRL principles in high-frequency trading suggests broader applicability to other domains requiring real-time decision making with complex action spaces. Future work will focus on extending the framework to more diverse financial instruments and developing online adaptation mechanisms for the frozen experts.

% \section*{Acknowledgments}

% We thank the anonymous reviewers for their constructive feedback. This work was supported by [funding information].

% Bibliography (use appropriate style)
\bibliographystyle{apalike}
%\biboptions{authoryear}
\bibliography{references}

\appendix

\section{Expert Market-Making Models}
\label{appendix:Expert Market-Making Models}

\subsection{Avellaneda-Stoikov (AS) Model}
The Avellaneda-Stoikov (AS) model~\cite{avellaneda2008high} optimizes bid and ask spreads by maximizing the expected utility of terminal wealth. Taking the coefficients of drift, jump intensity,  self-excitation and mutual-excitation to be zero in above settings , The value function $u(s, x, q, t)$ is governed by the Hamilton-Jacobi-Bellman (HJB) equation (details in Appendix):

\begin{align}
\frac{\partial u}{\partial t} + \frac{1}{2} \sigma^2 \frac{\partial^2 u}{\partial s^2} 
&+ \max_{\delta_b} \lambda_b\left(\delta_b\right)\left[u\left(s, x-s+\delta_b, q+1, t\right)-u(s, x, q, t)\right] \notag \\
&+ \max_{\delta_a} \lambda_a\left(\delta_a\right)\left[u\left(s, x+s+\delta_a, q-1, t\right)-u(s, x, q, t)\right] = 0, \label{eq:HJB_AS}
\end{align}

with terminal condition $u(s, x, q, T) = -e^{-\gamma\left(x + q S_T\right)}$.

The optimal bid and ask spreads are:

\begin{align}
\delta_{\text{bid}}^{\text{AS}} &= \frac{\gamma \sigma^2 \tau}{2} + \frac{1}{\gamma} \ln\left(1 + \frac{\gamma}{k}\right) - q \gamma \sigma^2 \tau, \label{eq:delta_bid_AS} \\
\delta_{\text{ask}}^{\text{AS}} &= \frac{\gamma \sigma^2 \tau}{2} + \frac{1}{\gamma} \ln\left(1 + \frac{\gamma}{k}\right) + q \gamma \sigma^2 \tau, \label{eq:delta_ask_AS}
\end{align}

% where:

% \begin{itemize}
%     \item $\gamma$ is the risk aversion parameter.
%     \item $\sigma$ is the asset's volatility.
%     \item $\tau = T - t$ denotes the remaining time horizon.
%     \item $k$ measures the sensitivity of order arrivals to spreads.
%     \item $q$ is the current inventory level.
% \end{itemize}

In the model, $\gamma$ is defined as the risk aversion parameter, which quantifies the trader's preference for avoiding risk relative to seeking returns. The variable $\sigma$ represents the asset's volatility, indicating the degree of price fluctuation over time. The term $\tau = T - t$ denotes the remaining time horizon, calculating the time left until the end of the trading period. The parameter $k$ measures the sensitivity of order arrivals to price spreads, demonstrating how spread changes influence trading activity. Lastly, $q$ is the current inventory level, reflecting the amount of the asset held by the trader.

\subsection{Guéant-Lehalle-Fernandez-Tapia (GLFT) Model}
The Guéant-Lehalle-Fernandez-Tapia (GLFT) model~\cite{gueant2013dealing} offers an approximate solution for optimal spreads by incorporating additional terms into the HJB equation. The assumptions are keeping the same as AS model expect. There is no limitation to the terminal time T. The optimal spreads are:

\begin{align}
    c_1 &= \frac{1}{\gamma} \ln\left(1 + \frac{\gamma}{k}\right), \\
    c_2 &= \sqrt{\frac{\gamma}{2 A k} \left(1 + \frac{\gamma}{k}\right)^{\left(\frac{k}{\gamma} + 1\right)}}, \\
    \delta_{\text{bid}}^{\text{GLFT}} &= c_1 + \frac{\sigma c_2}{2} + \sigma c_2 q, \\
    \delta_{\text{ask}}^{\text{GLFT}} &= c_1 + \frac{\sigma c_2}{2} - \sigma c_2 q, \\
\end{align}

% where:

% \begin{itemize}

%     \item $A$ denotes the base arrival rate of orders.
%     \item $c_1$ and $c_2$ are coefficients that adjust the bid and ask spreads based on risk aversion, volatility, and time horizon.
%     \item $q$ represents the current inventory level.
% \end{itemize}

In the model, $A$ denotes the base arrival rate of orders. The coefficients $c_1$ and $c_2$ are used to adjust the bid and ask spreads based on factors such as risk aversion, volatility, and time horizon. Lastly, $q$ represents the current inventory level, indicating the amount of the asset that is held.

\subsection{Modified Guéant-Lehalle-Fernandez-Tapia (GLFT) Model with Drift}
Keeping everything unchanged except for the drift part, the new modified GLFT solution which hasn't been shown in the original paper\cite{gueant2013dealing}.
\[|q| = Q_{\max}\]

\[dS_t = \mu dt + \sigma dW_t\]

\[(\delta^b_{\infty})^* = \frac{1}{\gamma}\ln(1 + \frac{\gamma}{k}) + \left[-\frac{\mu}{\gamma\sigma^2} + \frac{2q+1}{2}\right]\sqrt{\frac{\sigma^2}{2kA}(1 + \frac{\gamma}{k})^{1+\frac{k}{\gamma}}}\]

\[(\delta^a_{\infty})^* = \frac{1}{\gamma}\ln(1 + \frac{\gamma}{k}) + \left[\frac{\mu}{\gamma\sigma^2} - \frac{2q-1}{2}\right]\sqrt{\frac{\sigma^2}{2kA}(1 + \frac{\gamma}{k})^{1+\frac{k}{\gamma}}}\]

\subsection{Proximal Policy Optimization (PPO) for High-Frequency Market Making}
\label{sec:PPO}

To apply Proximal Policy Optimization (PPO) in a high-frequency trading (HFT) environment, we formulate the trading process as a Markov Decision Process (MDP). The input to the policy model consists of high-dimensional market state features, such as limit order book snapshots, recent trade volumes, price trends, and microstructure indicators sampled at millisecond intervals. The action space can include discrete actions like placing a buy/sell/cancel order at a specific price level. The reward function is designed to capture profit-and-loss (PnL) adjusted for market impact and inventory risk, including components like realized spread, execution cost, and inventory penalties. The objective is to train a policy that maximizes expected cumulative rewards.

In the context of market making, the RL framework can be formalized as a Markov Decision Process (MDP) defined by the tuple $(\mathcal{S}, \mathcal{A}, \mathcal{P}, \mathcal{R}, \gamma)$, 
% where:
% \begin{itemize}
%     \item $\mathcal{S}$ is the state space.
%     \item $\mathcal{A}$ is the action space.
%     \item $\mathcal{P}$ represents the state transition probabilities.
%     \item $\mathcal{R}$ is the reward function.
%     \item $\gamma \in [0,1)$ is the discount factor.
% \end{itemize}
where, $S$ represents the state space, and $A$ denotes the action space. The state transition probabilities are symbolized by $P$, which dictates the likelihood of transitioning from one state to another given a particular action. $R$ is the reward function, which assigns a reward based on the state and action taken. Finally, $\gamma \in [0, 1]$ is the discount factor, used to determine the present value of future rewards, allowing for the evaluation of long-term versus immediate benefits.

For the space state, at each discrete time step $t$, the state $s_t$ encapsulates essential information required for decision-making:
\[
s_t = (t, x_t, q_t, S_t, B_{t-1}-A_{t-1}),
\]
% where:
% \begin{itemize}
%     \item $t$ is the timestamp.
%     \item $x_t$ denotes the current cash position.
%     \item $q_t$ represents the current inventory (number of shares held).
%     \item $S_t$ is the current underlying stock price.
%     \item $A_{t-1}-B_{t-1}$ is the difference of ask and bid quotes from the previous time step.
% \end{itemize}

In the model, $t$ is the timestamp indicating the specific point in time under consideration. The variable $x_t$ denotes the current cash position, while $q_t$ represents the current inventory, which refers to the number of shares held. The term $S_t$ is the current underlying stock price, providing a snapshot of the stock's market value at time $t$. $A_{t-1} - B_{t-1}$ is the difference between the ask and bid quotes from the previous time step, reflecting the spread.

For the action space, the agent's action $a_t$ consists of setting new ask and bid quotes relative to the mid-price $M_t$. Specifically:
\[
a_t = (\delta^{\text{ask}}_t, \delta^{\text{bid}}_t),
\]
% where:
% \begin{itemize}
%     \item $\delta^{\text{ask}}_t$ is the offset of the ask quote from the mid-price, setting the ask price to $M_t + \delta^{\text{ask}}_t$.
%     \item $\delta^{\text{bid}}_t$ is the offset of the bid quote from the mid-price, setting the bid price to $M_t - \delta^{\text{bid}}_t$.
% \end{itemize}

In the model, $\delta_{\text{ask}t}$ represents the offset of the ask quote from the mid-price, which sets the ask price to $M_t + \delta{\text{ask}t}$. Conversely, $\delta{\text{bid}t}$ is the offset of the bid quote from the mid-price, setting the bid price to $M_t - \delta{\text{bid}_t}$.

For the reward function, it is designed to balance profitability from the bid-ask spread against the risk associated with inventory holding. The formulation is:
\[
r_t = \text{PnL}_t - \lambda q_t^2,
\]
% where:
% \begin{itemize}
%     \item $\text{PnL}_t$ represents the profit and loss at time $t$, derived from executed trades.
%     \item $\lambda$ is a risk aversion parameter penalizing large inventory positions.
% \end{itemize}
$PnL_t$ represents the profit and loss at time $t$, derived from executed trades. The parameter $\lambda$ is a risk aversion parameter that penalizes large inventory positions.

After defining the MDP, policy gradient methods is used to optimize the policy $\pi_\theta(a|s)$ parameterized by $\theta$ to maximize the expected cumulative reward:
\[
J(\theta) = \mathbb{E}_{\tau \sim \pi_\theta} \left[ \sum_{t=0}^{T} \gamma^t r_t \right],
\]
where $\tau = (s_0, a_0, s_1, a_1, \ldots, s_T)$ denotes a trajectory sampled from the policy.
The output of the policy network is a parameterized probability distribution over actions, from which the actual trading action is sampled.

\section{Market Environment}
\label{appendix: arameters for Test Environemnt}

\subsection{Gym Environment Implementation}

We implement the market making environment with Hawkes process-driven order arrivals following OpenAI Gym standards:

\begin{algorithm}[H]
\caption{MDP Simulation of \texttt{HawkesMarketMakingEnv}}
\label{alg:hawkes-mdp}
\begin{algorithmic}[1]
\REQUIRE Environment parameters $(T,\Delta t,\mu,\sigma,\gamma,S_0,q_{\max},\lambda_J,\mu_J,\sigma_J,\mu_b,\mu_a)$ where $\gamma$: risk-aversion; 
\STATE $N \gets T/\Delta t$

\STATE \textbf{Function Reset():}
\STATE \hspace{1em} $t_0\gets 0$, $q_0\gets 0$, $S_0\gets S_0$, $X_0\gets 1000$
\STATE \hspace{1em} $\mathcal{N}_b\gets \emptyset$, $\mathcal{N}_a\gets \emptyset$ \COMMENT{Buy/sell order arrival times}
\STATE \hspace{1em} Simulate price process $\{S_i\}_{i=0}^N$ using Brownian motion with jumps
\STATE \hspace{1em} \textbf{return} $\mathbf{s}_0 = (t_0,X_0,q_0,S_0,0)$

\STATE \textbf{Function Step}($\mathbf{a}_t = (\delta^b_t, \delta^a_t) \in \mathbb{R}_+^2$): \COMMENT{bid/ask spreads}
\STATE $W_t \gets X_t + q_t \cdot S_t$ \COMMENT{wealth before action}
\STATE $P^b_t \gets S_t - \delta^b_t$ \COMMENT{bid price}
\STATE $P^a_t \gets S_t + \delta^a_t$ \COMMENT{ask price}

\STATE \textbf{Update Hawkes intensities:}
\STATE $\lambda^b_t \gets \mu_b + \sum_{t_i \in \mathcal{N}_b} \alpha_{bb} e^{-\beta(t-t_i)} + \sum_{t_j \in \mathcal{N}_a} \alpha_{ba} e^{-\beta(t-t_j)}$
\STATE $\lambda^a_t \gets \mu_a + \sum_{t_i \in \mathcal{N}_a} \alpha_{aa} e^{-\beta(t-t_i)} + \sum_{t_j \in \mathcal{N}_b} \alpha_{ab} e^{-\beta(t-t_j)}$

\STATE \textbf{Adjust intensities for spread effect:}
\STATE $\tilde{\lambda}^b_t \gets \lambda^b_t \cdot e^{-k\delta^b_t}$
\STATE $\tilde{\lambda}^a_t \gets \lambda^a_t \cdot e^{-k\delta^a_t}$

\STATE \textbf{Simulate order arrivals:}
\STATE $p^b_t \gets 1 - e^{-\tilde{\lambda}^b_t \Delta t}$, $p^a_t \gets 1 - e^{-\tilde{\lambda}^a_t \Delta t}$
\STATE Draw $U_b, U_a \sim \text{Uniform}(0,1)$
\STATE $\mathbb{1}_{\text{bid}} \gets (U_b < p^b_t)$, $\mathbb{1}_{\text{ask}} \gets (U_a < p^a_t)$

\IF{$\mathbb{1}_{\text{bid}}$}
\STATE $q_{t+\Delta t} \gets q_t + 1$, $X_{t+\Delta t} \gets X_t - P^b_t$
\STATE $\mathcal{N}_b \gets \mathcal{N}_b \cup \{t+\Delta t\}$
\ELSE
\STATE $q_{t+\Delta t} \gets q_t$, $X_{t+\Delta t} \gets X_t$
\ENDIF

\IF{$\mathbb{1}_{\text{ask}}$}
\STATE $q_{t+\Delta t} \gets q_{t+\Delta t} - 1$, $X_{t+\Delta t} \gets X_{t+\Delta t} + P^a_t$
\STATE $\mathcal{N}_a \gets \mathcal{N}_a \cup \{t+\Delta t\}$
\ENDIF

\STATE $t \gets t+\Delta t$
\STATE $S_{t+\Delta t} \gets S_{\text{step}+1}$ \COMMENT{from pre-simulated price process}
\STATE $\text{done}\gets(t\ge T)$
\STATE $W_{t+\Delta t} \gets X_{t+\Delta t} + q_{t+\Delta t} \cdot S_{t+\Delta t}$
\STATE $\text{reward}\gets W_{t+\Delta t} - W_t$ \COMMENT{PnL change}
\STATE \textbf{return} $(t,X_{t+\Delta t},q_{t+\Delta t},S_{t+\Delta t},P^a_t-P^b_t)$, $\text{reward}$, $\text{done}$, $\text{info}$

\end{algorithmic}
\end{algorithm}

\begin{algorithm}[H]
\caption{Price Process Simulation with Jumps and Brownian Motion}
\label{alg:price-simulation}
\begin{algorithmic}[1]
\REQUIRE Parameters $(S_0,\mu,\sigma,\lambda_J,\mu_J,\sigma_J,T,N)$
\STATE $\Delta t \gets T/N$
\STATE $S_0 \gets S_0$
\FOR{$i = 1$ to $N$}
\STATE Draw $U \sim \text{Uniform}(0,1)$
\IF{$U < \lambda_J \Delta t$}
\STATE Draw $J \sim \mathcal{N}(\mu_J, \sigma_J^2)$ \COMMENT{Jump size}
\STATE $S_i \gets S_{i-1} \cdot e^{J}$
\ELSE
\STATE Draw $Z \sim \mathcal{N}(0,1)$ \COMMENT{Brownian increment}
\STATE $dS \gets (\mu - \frac{1}{2}\sigma^2)\Delta t + \sigma\sqrt{\Delta t} \cdot Z$
\STATE $S_i \gets S_{i-1} \cdot e^{dS}$
\ENDIF
\ENDFOR
\STATE \textbf{return} $\{S_i\}_{i=0}^N$
\end{algorithmic}
\end{algorithm}

\begin{table}[H]
    \centering
    \caption{MDP Components of \texttt{HawkesMarketMakingEnv}}
    \label{tab:hawkes-mdp}
    \renewcommand{\arraystretch}{1.5}
    \small
    \begin{tabular}{@{}p{2.8cm}p{7.5cm}p{4.2cm}@{}}
        \toprule
        \textbf{Element} & \textbf{Definition} & \textbf{Source-code reference} \\ 
        \midrule
        
        \textbf{State} $\mathbf{s}_t$ &
        Tuple: $\bigl(\text{time},\;\text{cash},\;\text{inventory},\;\text{mid-price},\;\text{spread}\bigr)$
        & \texttt{\_get\_obs()} \\
        \addlinespace[4pt]
        
        \textbf{Action} $\mathbf{a}_t$ &
        Continuous vector in $\mathbb{R}_+^2$: $(\delta^b_t, \delta^a_t)$ representing bid and ask spreads
        & \texttt{step()} \\
        \addlinespace[4pt]
        
        \textbf{Dynamics} &
        \begin{minipage}[t]{7.3cm}
        \vspace{1pt}
        $\bullet$ Jump-diffusion with Brownian motion for asset price\\
        $\bullet$ Hawkes process for order arrivals (self/cross-excitation)\\
        $\bullet$ Spread-dependent intensity modulation: $\lambda e^{-k\delta}$\\
        $\bullet$ Inventory updates: $+1$ (buy), $-1$ (sell)
        \vspace{1pt}
        \end{minipage}
        & \texttt{\_update\_hawkes\_intensity()}, \texttt{step()} \\
        \addlinespace[4pt]
        
        \textbf{Reward} &
        PnL change: $r_t = (X_{t+\Delta t} + q_{t+\Delta t}S_{t+\Delta t}) - (X_t + q_tS_t)$
        & reward computation \\
        \addlinespace[4pt]
        
        \textbf{Termination} &
        End of horizon $t = T$
        & \texttt{done} flag \\ 
        
        \bottomrule
    \end{tabular}
\end{table}

\textbf{Key Environment Features:}
\begin{itemize}
\item Time discretization: $N = T/\Delta t$ steps (default: 100 steps over $T = 1.0$)
\item Hawkes process parameters: 
    \begin{itemize}
    \item Self-excitation: $\alpha_{bb} = \alpha_{aa} = 0.7$
    \item Cross-excitation: $\alpha_{ab} = \alpha_{ba} = 0.3$
    \item Decay rate: $\beta = 0.1$
    \end{itemize}
\item Price dynamics: Geometric Brownian motion with compound Poisson jumps
\item Order arrival intensity adjustment: $\lambda(\delta) = \lambda_0 \cdot e^{-k\delta}$ where $k$ is the spread sensitivity parameter
\item No terminal constraint on inventory (unlike optimal execution)
\end{itemize}

\textbf{Optimal Spread Strategies:}

% \begin{table}[H]
%     \centering
%     \caption{Implemented Optimal Spread Strategies}
%     \label{tab:optimal-strategies}
%     \renewcommand{\arraystretch}{1.5}
%     \small
%     \begin{tabular}{@{}p{2cm}p{10cm}@{}}
%         \toprule
%         \textbf{Strategy} & \textbf{Formula} \\ 
%         \midrule
        
%         \textbf{AS Model} &
%         \begin{minipage}[t]{10cm}
%         \vspace{1pt}
%         $\delta^b = \frac{\gamma\sigma^2\tau}{2} + \frac{1}{\gamma}\ln\left(1+\frac{\gamma}{k}\right) - q\gamma\sigma^2\tau$\\
%         $\delta^a = \frac{\gamma\sigma^2\tau}{2} + \frac{1}{\gamma}\ln\left(1+\frac{\gamma}{k}\right) + q\gamma\sigma^2\tau$\\
%         where $\tau = T - t$ is the remaining time
%         \vspace{1pt}
%         \end{minipage} \\
%         \addlinespace[4pt]
        
%         \textbf{GLFT Model} &
%         \begin{minipage}[t]{10cm}
%         \vspace{1pt}
%         $\delta^b = c_1 + \frac{\Delta}{2}\sigma c_2 + q\sigma c_2$\\
%         $\delta^a = c_1 + \frac{\Delta}{2}\sigma c_2 - q\sigma c_2$\\
%         where $\Delta = T - t$,\\
%         $c_1 = \frac{1}{\gamma\Delta}\ln\left(1+\frac{\gamma\Delta}{k}\right)$,\\
%         $c_2 = \sqrt{\frac{\gamma}{2A\Delta k}\left(1+\frac{\gamma\Delta}{k}\right)^{\frac{k}{\gamma\Delta}+1}}$
%         \vspace{1pt}
%         \end{minipage} \\
        
%         \bottomrule
%     \end{tabular}
% \end{table}

We evaluate FinFlowRL on a simulated limit order book based on Hawkes processes, which capture the self-exciting nature of order arrivals. The environment parameters include:

\begin{itemize}
\item Volatility: $\sigma \in \{0.02, 0.25\}$ (Low/High)
\item Order arrival intensity: $\lambda \in \{25, 50\}$ (Low/High)
\item Hurst exponent: $H = 0.5$ (Brownian motion)
\item Drift rate: $\mu = 0$ (no systematic drift)
\item Episode length: 100 time steps
\item Inventory constraints: $Q_{max} = 10$
\item Inventory penalty: $\gamma = 0.1$
\end{itemize}

This creates four market modes: 
\begin{itemize}
\item HH: High volatility ($\sigma=0.25$), High demand ($\lambda=50$)
\item HL: High volatility ($\sigma=0.25$), Low demand ($\lambda=25$)
\item LH: Low volatility ($\sigma=0.02$), High demand ($\lambda=50$)
\item LL: Low volatility ($\sigma=0.02$), Low demand ($\lambda=25$)
\end{itemize}

\subsection{Implementation Details}

\begin{itemize}
\item \textbf{MeanFlow Expert}: UNet architecture with 2.1M parameters, pre-trained on expert demonstrations
\item \textbf{Noise Policy}: 3-layer MLP with 256 hidden units, outputting $(T_{pred} \times d_{act})$ dimensional noise
\item \textbf{Critic}: 3-layer MLP with 256 hidden units
\item \textbf{Training}: 256 parallel environments, batch size 25,600, learning rate $5 \times 10^{-5}$
\item \textbf{Horizons}: $T_{obs} = 2$, $T_{pred} = 16$, $T_{exec} = 8$
\end{itemize}

\section{Calculation of Evaluation Metrics}
\label{Calculation of Evaluation Metrics}
\subsection*{1. Cumulative Return (CR)}
The Cumulative Return measures the total percentage change in the value of an investment over a specific period.

% \subsubsection*{Method 1: Using Initial and Final Values}
% If $V_{\text{initial}}$ is the initial value of the investment and $V_{\text{final}}$ is the final value of the investment:
% \[
% \text{CR} = \frac{V_{\text{final}} - V_{\text{initial}}}{V_{\text{initial}}} \times 100\%
% \]

% \subsubsection*{Method 2: Using a series of periodic returns}
If $r_i$ is the return for period $i$, and there are $n$ periods:
\[
\text{CR} = \left( \prod_{i=1}^{n} (1 + r_i) - 1 \right) \times 100\%
\]
where:
\begin{itemize}
    \item $r_i$: Return for period $i$.
    \item $n$: Total number of periods.
\end{itemize}

\subsection*{2. Sharpe Ratio (SR)}
The Sharpe Ratio measures the risk-adjusted return of an investment or strategy. It quantifies the average return earned in excess of the risk-free rate per unit of volatility or total risk.
\[
\text{SR} = \frac{\mathbb{E}[R_p - R_f]}{\sigma_{p}}
\]

where:
\begin{itemize}
    \item $\mathbb{E}[R_p - R_f]$: The expected value of the excess return of the portfolio over the risk-free rate.
    \item $\bar{R}_p$: The average (sample mean) return of the portfolio or investment over a period.
    \item $\bar{R}_f$: The average (sample mean) risk-free rate of return over the same period (e.g., yield on short-term government bonds).
    \item $\sigma_p$: The standard deviation of the portfolio's excess returns ($R_p - R_f$). (Sometimes, the standard deviation of the portfolio's returns, $\sigma(R_p)$, is used as an approximation, especially if $R_f$ is constant or has very low volatility).
\end{itemize}

\subsection*{3. Maximum Drawdown (MDD)}
The Maximum Drawdown is the largest percentage drop in the value of an investment or portfolio from a previous peak to a subsequent trough over a specific period. Let $V(t)$ be the value of the portfolio at time $t$ over the period $[0, T]$.

First, define the running peak value up to time $t$ as $M(t)$:
\[
M(t) = \max_{0 \le \tau \le t} V(\tau)
\]
Then, the drawdown $D(t)$ at any time $t$ is:
\[
D(t) = \frac{M(t) - V(t)}{M(t)}
\]
The Maximum Drawdown (MDD) is the maximum value of $D(t)$ over the entire period:
\[
\text{MDD} = \max_{0 \le t \le T} D(t)
\]
This MDD is usually expressed as a positive percentage (e.g., an MDD of 0.25 means a 25\% maximum loss from a peak).

where:
\begin{itemize}
    \item $V(t)$: Value of the investment at time $t$.
    \item $T$: The end of the observation period.
    \item $\tau$: A time index variable ranging from $0$ to $t$.
\end{itemize}

\end{document}